\documentclass[12pt,a4paper]{article}

\usepackage{amsmath}

\usepackage{epsfig}

\usepackage{rotating} %

\usepackage{dcolumn}

\usepackage{url}

\usepackage{hyperref}

\usepackage{ifthen} 
\usepackage{subfigure} 
\usepackage{rotating}

\usepackage{lineno}  
\usepackage{graphicx}  

\usepackage{xspace}  
\usepackage{color} %
\usepackage{colortbl} %
\usepackage{ctable}
\usepackage{placeins}
\usepackage{multirow}

\newboolean{articletitles}
\setboolean{articletitles}{true} 

\newboolean{pdflatex}
\setboolean{pdflatex}{true} 
\newboolean{uprightparticles}
\setboolean{uprightparticles}{false} 

\usepackage{amssymb}
\usepackage{amsfonts}
\usepackage{upgreek}
\usepackage{rotating}
\usepackage{cite}
\usepackage{graphicx}

\hypersetup{ pdfborder=0 0 0, urlbordercolor=1 0 0 }

\ifthenelse{\boolean{uprightparticles}}%
{

 \def\Pmu         {\ensuremath{\upmu}\xspace}

 \def\Ppsi        {\ensuremath{\uppsi}\xspace}

 \def\PDelta      {\ensuremath{\Delta}\xspace}                 
 \def\PXi      {\ensuremath{\Xi}\xspace}                 
 \def\PLambda      {\ensuremath{\Lambda}\xspace}                 
 \def\PSigma      {\ensuremath{\Sigma}\xspace}                 
 \def\POmega      {\ensuremath{\Omega}\xspace}                 
 \def\PUpsilon      {\ensuremath{\Upsilon}\xspace}                 
 
 \def\PB      {\ensuremath{\mathrm{B}}\xspace}                 
                  
 \def\PD      {\ensuremath{\mathrm{D}}\xspace}

 \def\PJ      {\ensuremath{\mathrm{J}}\xspace}                 
 \def\PK      {\ensuremath{\mathrm{K}}\xspace}

 \def\Pi      {\ensuremath{\mathrm{i}}\xspace}

}
{

 \def\Pmu         {\ensuremath{\mu}\xspace}

 \def\Ppsi        {\ensuremath{\psi}\xspace}                 
                  
 \mathchardef\PDelta="7101
 \mathchardef\PXi="7104
 \mathchardef\PLambda="7103
 \mathchardef\PSigma="7106
 \mathchardef\POmega="710A
 \mathchardef\PUpsilon="7107
                  
 \def\PB      {\ensuremath{B}\xspace}                 
                  
 \def\PD      {\ensuremath{D}\xspace}

 \def\PJ      {\ensuremath{J}\xspace}                 
 \def\PK      {\ensuremath{K}\xspace}

 \def\Pi      {\ensuremath{i}\xspace}

}

\def\mup        {\ensuremath{\Pmu^+}\xspace}
\def\mun        {\ensuremath{\Pmu^-}\xspace} 
\def\mumu       {\ensuremath{\Pmu^+\Pmu^-}\xspace}

\def\kaon  {\ensuremath{\PK}\xspace}
  \def\Kbar  {\kern 0.2em\overline{\kern -0.2em \PK}{}\xspace}

\def\Kz    {\ensuremath{\kaon^0}\xspace}
\def\Kzb   {\ensuremath{\Kbar^0}\xspace}
\def\KzKzb {\ensuremath{\Kz \kern -0.16em \Kzb}\xspace}
\def\Kp    {\ensuremath{\kaon^+}\xspace}
\def\Km    {\ensuremath{\kaon^-}\xspace}

\def\KpKm  {\ensuremath{\Kp \kern -0.16em \Km}\xspace}

  \def\Dbar    {\kern 0.2em\overline{\kern -0.2em \PD}{}\xspace}
\def\D       {\ensuremath{\PD}\xspace}

\def\Dz      {\ensuremath{\D^0}\xspace}
\def\Dzb     {\ensuremath{\Dbar^0}\xspace}
\def\DzDzb   {\ensuremath{\Dz {\kern -0.16em \Dzb}}\xspace}
\def\Dp      {\ensuremath{\D^+}\xspace}
\def\Dm      {\ensuremath{\D^-}\xspace}

\def\DpDm    {\ensuremath{\Dp {\kern -0.16em \Dm}}\xspace}

\def\B       {\ensuremath{\PB}\xspace}
  \def\Bbar    {\kern 0.18em\overline{\kern -0.18em \PB}{}\xspace}

\def\Bu      {\ensuremath{\B^+}\xspace}

\def\Bd      {\ensuremath{\B^0}\xspace}
\def\Bs      {\ensuremath{\B^0_s}\xspace}

\def\jpsi     {\ensuremath{{\PJ\mskip -3mu/\mskip -2mu\Ppsi\mskip 2mu}}\xspace}
\def\psitwos  {\ensuremath{\Ppsi{(2S)}}\xspace}

  \def\Y#1S{\ensuremath{\PUpsilon{(#1S)}}\xspace}
\def\OneS  {\Y1S}
\def\TwoS  {\Y2S}
\def\ThreeS{\Y3S}

\newcommand{\mBd}{\ensuremath{m_{\Bd}}\xspace}

\newcommand{\mBs}{\ensuremath{m_{\Bs}}\xspace}

\newcommand{\tev}{\ensuremath{\mathrm{\,Te\kern -0.1em V}}\xspace}
\newcommand{\gev}{\ensuremath{\mathrm{\,Ge\kern -0.1em V}}\xspace}
\newcommand{\mev}{\ensuremath{\mathrm{\,Me\kern -0.1em V}}\xspace}
\newcommand{\kev}{\ensuremath{\mathrm{\,ke\kern -0.1em V}}\xspace}
\newcommand{\ev}{\ensuremath{\mathrm{\,e\kern -0.1em V}}\xspace}
\newcommand{\gevc}{\ensuremath{{\mathrm{\,Ge\kern -0.1em V\!/}c}}\xspace}
\newcommand{\mevc}{\ensuremath{{\mathrm{\,Me\kern -0.1em V\!/}c}}\xspace}
\newcommand{\gevcc}{\ensuremath{{\mathrm{\,Ge\kern -0.1em V\!/}c^2}}\xspace}
\newcommand{\gevgevcccc}{\ensuremath{{\mathrm{\,Ge\kern -0.1em V^2\!/}c^4}}\xspace}
\newcommand{\mevcc}{\ensuremath{{\mathrm{\,Me\kern -0.1em V\!/}c^2}}\xspace}

\def\invfb   {\ensuremath{\mbox{\,fb}^{-1}}\xspace}

\def\BR         {{\ensuremath{\cal B}\xspace}}

\newcommand{\decay}[2]{\ensuremath{#1\!\to #2}\xspace}         

\def\to                 {\ensuremath{\rightarrow}\xspace}

\def\CP                {\ensuremath{C\!P}\xspace}

\def\gsim{{~\raise.15em\hbox{$>$}\kern-.85em
          \lower.35em\hbox{$\sim$}~}\xspace}
\def\lsim{{~\raise.15em\hbox{$<$}\kern-.85em
          \lower.35em\hbox{$\sim$}~}\xspace}

\def\BsToJPsiPhi  {\decay{\Bs}{\jpsi\phi}\xspace}

\def\AT#1     {\ensuremath{A_T^{#1}}\xspace}           

\def\Bsmm     {\decay{\Bs}{\mup\mun}\xspace}
\def\Bdmm     {\decay{\Bd}{\mup\mun}\xspace}

\def\C#1      {\ensuremath{\mathcal{C}_{#1}}}                       
\def\Cp#1     {\ensuremath{\mathcal{C}_{#1}^{'}}}                    
\def\Ceff#1   {\ensuremath{\mathcal{C}_{#1}^{\mathrm{(eff)}}}}        
\def\Cpeff#1  {\ensuremath{\mathcal{C}_{#1}^{'\mathrm{(eff)}}}}       
\def\Ope#1    {\ensuremath{\mathcal{O}_{#1}}}                       
\def\Opep#1   {\ensuremath{\mathcal{O}_{#1}^{'}}}                    

\def\BuToJPsiK {\decay{\Bu}{\jpsi\Kp}}

\newcommand{\CL}{CL\xspace}
\newcommand{\CLsb}{\ensuremath{\textrm{CL}_{\textrm{s+b}}}\xspace}
\newcommand{\CLs}{\ensuremath{\textrm{CL}_{\textrm{s}}}\xspace}
\newcommand{\CLb}{\ensuremath{\textrm{CL}_{\textrm{b}}}\xspace}
\newcommand{\BF}{branching fraction\xspace}

\newcommand{\bbdim}{\ensuremath{b\bar{b}\to \mu^+ \mu^- X}\xspace}

\newcommand{\Bsmumu}{\ensuremath{\Bs\to\mu^+\mu^-}\xspace}
\newcommand{\Bdmumu}{\ensuremath{\Bd\to\mu^+\mu^-}\xspace}

\newcommand{\BsKK}{\ensuremath{\Bs\to K^+K^-}\xspace}

\newcommand{\BdKpi}{\ensuremath{\Bd\to K^+\pi^-}\xspace}

\newcommand{\Bhh}{\ensuremath{B^0_{(s)}\to h^+h^{'-}}\xspace}
\newcommand{\Bmm}{\ensuremath{B^0_{(s)}\to \mu^+\mu^-}\xspace}

\newcommand{\BuJpsiK}{\ensuremath{B^+\to J/\psi K^+}\xspace}
\newcommand{\BuJpsimumuK}{\ensuremath{B^+\to J/\psi(\mu^+\mu^-)K^+}\xspace}

\newcommand{\Jpsi}{\ensuremath{J/\psi}\xspace}

\newcommand{\BsJpsiPhi}{\ensuremath{B^0_s\to J/\psi \phi}\xspace}

\newcommand{\BRof}[1]{\ensuremath{{\cal B}(#1)}\xspace}

\newcommand{\MeVcc}{\ensuremath{\,{\rm MeV}/c^2}\xspace}
\newcommand{\IP}{\ensuremath{{\rm IP}}\xspace}
\newcommand{\gl}{\ensuremath{{\rm GL}}\xspace}

\newcommand\TTstrut{\rule{0pt}{3.2ex}}

\newcommand\BBstrut{\rule[-1.8ex]{0pt}{0pt}}

\newcommand{\tabcaption}[1]{  
\vspace{-\abovecaptionskip} %
\caption[#1]{#1}
\vspace{2mm}
}

\usepackage{mciteplus}

\begin{document}

\begin{titlepage}

\belowpdfbookmark{Title page}{title}

\pagenumbering{roman}

\vspace*{-1.5cm}

\centerline{\large EUROPEAN ORGANIZATION FOR NUCLEAR RESEARCH (CERN)}

\vspace*{1.5cm}

\hspace*{-5mm}\begin{tabular*}{16cm}{lc@{\extracolsep{\fill}}r}

\vspace*{-12mm}\mbox{\!\!\!\epsfig{figure=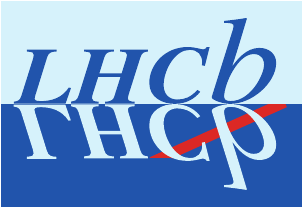,width=.12\textwidth}}& & \\

&& CERN-PH-EP-2011-186\\
&& LHCb-PAPER-2011-025\\

\end{tabular*}

\vspace*{4cm}

\begin{center}

{\bf\huge\boldmath {Search for the rare decays \Bsmumu and \Bdmumu } \\
}

\vspace*{2cm}

\normalsize {

The LHCb Collaboration\footnote{Authors are listed on the following pages.}

}

\end{center}

\vspace{\fill}

\centerline{\bf Abstract}

\vspace*{5mm}

\noindent
A search for the decays \Bsmumu and \Bdmumu is performed with 
0.37 fb$^{-1}$ of $pp$ collisions at $\sqrt{s}$ = 7~TeV
collected by the LHCb experiment in 2011. 
The upper limits on the branching fractions are \BRof \Bsmumu $< 1.6 \times 10^{-8}$ 
and \BRof \Bdmumu $<3.6 \times 10^{-9}$ at 95 \% confidence level.
A combination of these results with the LHCb limits obtained with the 2010 dataset leads to 
\BRof \Bsmumu $<1.4 \times 10^{-8}$ and \BRof \Bdmumu $<3.2 \times 10^{-9}$ at 95\,\% 
confidence level. 

\vspace*{1.cm}



\vspace{\fill}

\vspace*{0.5cm}

\end{titlepage}

\setcounter{page}{2}

\belowpdfbookmark{LHCb author list}{authors}

\noindent 
R.~Aaij$^{23}$, 
C.~Abellan~Beteta$^{35,n}$, 
B.~Adeva$^{36}$, 
M.~Adinolfi$^{42}$, 
C.~Adrover$^{6}$, 
A.~Affolder$^{48}$, 
Z.~Ajaltouni$^{5}$, 
J.~Albrecht$^{37}$, 
F.~Alessio$^{37}$, 
M.~Alexander$^{47}$
% ...

\noindent
{\footnotesize \it 
$ ^{1}$Centro Brasileiro de Pesquisas F\'{i}sicas (CBPF), Rio de Janeiro, Brazil\\
% ...
$ ^{5}$Clermont Universit\'{e}, Universit\'{e} Blaise Pascal, CNRS/IN2P3, LPC, Clermont-Ferrand, France\\
$ ^{6}$CPPM, Aix-Marseille Universit\'{e}, CNRS/IN2P3, Marseille, France\\
% ...
$ ^{23}$Nikhef National Institute for Subatomic Physics, Amsterdam, The Netherlands\\
% ...
$ ^{35}$Universitat de Barcelona, Barcelona, Spain\\
$ ^{36}$Universidad de Santiago de Compostela, Santiago de Compostela, Spain\\
$ ^{37}$European Organization for Nuclear Research (CERN), Geneva, Switzerland\\
% ...
$ ^{42}$H.H. Wills Physics Laboratory, University of Bristol, Bristol, United Kingdom\\
% ...
$ ^{47}$School of Physics and Astronomy, University of Glasgow, Glasgow, United Kingdom\\
$ ^{48}$Oliver Lodge Laboratory, University of Liverpool, Liverpool, United Kingdom\\
% ...
$ ^{n}$LIFAELS, La Salle, Universitat Ramon Llull, Barcelona, Spain\\
% ...
}

\cleardoublepage

\pagestyle{plain} 
\setcounter{page}{1}
\pagenumbering{arabic}

\section{Introduction}
\label{sec:intro}

Measurements of low-energy processes can provide indirect constraints
on particles that are too heavy to be produced directly.
This is particularly true for Flavour Changing Neutral Current
(FCNC) processes which are highly suppressed in the Standard Model (SM) 
and can only occur through higher-order diagrams. 
The SM predictions for the branching fractions  
of the FCNC decays\footnote{Inclusion of charged conjugated processes is implied
throughout.} \Bsmumu 
and \Bdmumu are \BRof \Bsmumu = $(3.2 \pm
0.2) \times 10^{-9}$ and \BRof \Bdmumu = $(0.10 \pm 0.01) \times 10^{-9}$~\cite{Buras2010mh,*Buras:2010wr}.
However, contributions from new processes or new heavy particles can significantly enhance these values.
For example, within Minimal Supersymmetric extensions of the SM (MSSM), 
in the large $\tan \beta$ regime,
\BRof \Bsmumu is found to be approximately proportional
to $\tan^6\beta$ \cite{Choudhury:1998ze,*Hamzaoui:1998nu,*Babu:1999hn,*Hall:1993gn,*Huang:1998vb}, where $\tan\beta$ is the ratio of the vacuum
expectation values of the two neutral \CP-even Higgs fields.
The branching fractions could therefore be enhanced by orders of magnitude 
for large values of $\tan \beta$.

The best published limits from the Tevatron 
are  \BRof \Bsmumu $<~5.1~\times~10^{-8}$ at 95\% confidence level (CL) 
by the D0 collaboration using 6.1 fb$^{-1}$ of data~\cite{d0_PLB}, 
and \BRof \Bdmumu$<6.0 \times 10^{-9}$ at 95\% \CL by the CDF collaboration
using 6.9 fb$^{-1}$ of data ~\cite{cdf_prl}.
In the same dataset the CDF collaboration observes an excess 
of \Bsmumu candidates 
compatible with \BRof \Bsmumu = $(1.8^{+1.1}_{-0.9}) \times 10^{-8}$
and with an upper limit of \BRof \Bsmumu $< 4.0 \times 10^{-8}$ at 95\% \CL.
The CMS collaboration has recently published \BRof \Bsmumu $< 1.9 \times 10^{-8}$ at 95\% \CL and
\BRof \Bdmumu $< 4.6 \times 10^{-9}$ at 95\% \CL using 1.14 fb$^{-1}$ of data ~\cite{cms}.
The LHCb collaboration has published the limits \cite{LHCb_paper} 
\BRof \Bsmumu $<5.4 \times 10^{-8}$ and  \BRof \Bdmumu$<1.5 \times 10^{-8}$ at 95$\%$ \CL
based on about 37 pb$^{-1}$ of integrated luminosity collected in the 2010 run.

This Letter presents an analysis of the data recorded by LHCb in the first half of 2011 which 
correspond to an integrated luminosity of
$\sim$ 0.37 fb$^{-1}$. The results of this analysis are then combined with those
published from the 2010 dataset. 

\section{The LHCb detector}
\label{sec:lhcbdet}

The LHCb detector \cite{LHCbdetector} is 
a single-arm forward spectrometer designed to study production and decays of hadrons containing
$b$ or $c$ quarks. The detector consists of a vertex locator (VELO) providing precise locations
of primary $pp$ interaction vertices and detached vertices of long lived hadrons. 

The momenta of charged particles are determined using information from the VELO together
with the rest of the tracking system, composed of a large area silicon tracker located
before a warm dipole magnet with a bending power of $\sim$  4 Tm, 
and a combination of silicon strip detectors and straw drift chambers
located after the magnet. Two Ring Imaging Cherenkov (RICH) detectors are
used for charged hadron identification in the momentum range 2--100 \gevc. 
Photon, electron and hadron candidates are identified by 
electromagnetic and hadronic calorimeters.
A muon system of alternating layers of iron and drift chambers provides muon identification. The
two calorimeters and the muon system provide the energy and momentum information
to implement a first level (L0) hardware trigger. An additional trigger level (HLT) is software
based, and its algorithms are tuned to the experimental operating condition.

Events with a muon final states are triggered using two L0 trigger decisions:
the single-muon decision, which requires one muon candidate with
a transverse momentum $p_{\rm T}$ larger than 1.5 \gevc, and the di-muon decision, which requires
two muon candidates with transverse momenta $p_{{\rm T},1}$ and $p_{{\rm T},2}$ satisfying the
relation $\sqrt{p_{{\rm T},1} \cdot p_{{\rm T},2}} > 1.3$ \gevc. 

The single muon trigger decision in the second trigger level (HLT)
includes a cut on the impact parameter (\IP) with respect to the 
primary vertex, which allows for a lower $p_{\rm T}$ requirement 
($p_{\rm T}>1.0$ GeV/$c$, $\rm IP> 0.1$~mm). 
The di-muon trigger decision requires muon pairs of opposite charge 
with $p_{\rm T}>500$ MeV/c,  forming a common vertex 
and with an invariant mass $m_{\mu\mu} > 4.7 $ GeV/$c^2$.   
A second trigger decision, primarily to select \jpsi events, requires 
$2.97 < m_{\mu\mu} < 3.21$ GeV/$c^2$. 
The remaining region of the di-muon invariant mass range is also covered by trigger decisions that 
in addition require the di-muon secondary vertex to be well separated from the primary vertex.

Events with purely hadronic final states are triggered by the L0 trigger if there is a
calorimeter cluster with transverse energy $E_{\rm T} > 3.6$ GeV.
Other HLT trigger decisions select generic displaced vertices, providing high 
efficiency for purely hadronic decays.

\section{Analysis strategy}
\label{sec:strategy}

Assuming the branching fractions predicted by the SM, and using the $b \bar{b}$ cross-section measured
by LHCb in the pseudorapidity interval $2 < \eta < 6$ and integrated over all transverse momenta
of $\sigma_{b \overline{b}} = 75\pm14 \,\mu$b~\cite{bbxsection}, 
approximately $3.9$ 
$B^0_{s}\to \mu^+\mu^-$ and 0.4 $B^0 \to \mu^+\mu^-$  events
are expected to be triggered, reconstructed and selected in the analysed sample embedded in a large background.

The general structure of the analysis is based upon the one described in
Ref.~\cite{LHCb_paper}. 
First a very efficient selection removes 
the biggest amount of background while keeping most of the signal within the LHCb acceptance.
The number of observed events is compared to the number of expected 
signal and background events in bins of two independent variables,
the invariant mass and the output of a multi-variate discriminant. The discriminant is 
a Boosted Decision Tree (BDT) constructed using the TMVA package~\cite{tmva}.
It supersedes the Geometrical Likelihood (GL) used in the previous analysis
\cite{LHCb_paper} as it has been found more performant in discriminating between 
signal and background events in simulated samples.
No data were used in the choice of the multivariate discriminant in order not 
to bias the result.

The combination of variables entering the BDT discriminant is optimized using simulated events.
The probability for a signal or background
event to have a given value of the BDT output 
is obtained from data using \Bhh candidates (where $h^{(')}$ can be a pion or a kaon) 
as signal and sideband \Bmm candidates 
as background. 

The invariant mass line shape of the signals is described by a Crystal Ball function \cite{crystalball} 
with parameters extracted from data control samples.
The central values of the masses are obtained from  \BdKpi and \BsKK samples.
The \Bs and \Bd
mass resolutions are estimated by interpolating those 
obtained with di-muon 
resonances ($J/\psi, \psi(2S)$ and $\Upsilon(1S,2S,3S)$) and cross-checked with a fit
to the invariant mass distributions of both inclusive \Bhh decays and exclusive \BdKpi decays.
The central values of the masses and the mass resolution are used to define the
signal regions.

The number of expected signal events, for a given branching fraction hypothesis, is obtained 
by normalizing to channels of known branching fractions: 
\BuToJPsiK, \BsToJPsiPhi and $B^0 \to K^+ \pi^-$. These channels
are selected in a way as similar as possible to the signals
in order to minimize the systematic uncertainty related to the different phase space accessible to each final state.

The BDT output and invariant mass distributions for combinatorial background events  in the signal regions
are obtained using fits of the mass distribution of events in the mass sidebands in bins of the BDT output.

The two-dimensional space formed by the invariant mass  
and the BDT output is binned. For each bin we count the number of candidates
observed in the data, and compute the expected number of signal events  
and the expected number of background events. 
The binning is unchanged with respect to the 2010 analysis \cite{LHCb_paper}.
The compatibility of the observed distribution of events in
all bins with the distribution expected for a given branching fraction hypothesis 
is computed using the \CLs method~\cite{Read_02}, which allows a given 
hypothesis to be excluded at a given confidence level.

\section{Selection}
\label{sec:selection}

The \Bmm selections 
require two muon candidates of opposite charge. 
Tracks are required to be of good quality 
and to be displaced with respect to any primary vertex. 
The secondary vertex is required to be well fitted ($\chi^2/{\rm nDoF} < 9$)
and must be separated from the primary vertex in the forward
direction by a distance of flight significance ($L/\sigma(L)$) greater than 15.
When more than one primary vertex is
reconstructed, the one that gives the minimum impact
parameter significance for the candidate is chosen.  The reconstructed candidate has to point to this
primary vertex (${\rm IP}/\sigma({\rm IP})<5$).

Improvements have been made to the selection developed for 2010 data \cite{LHCb_paper}.
The RICH is used to identify kaons in the $B^0_s \to J/\psi \phi$ normalization 
channel and the Kullback-Leibler (KL) distance  \cite{KLdistance,*KLdistance2} is used
to suppress duplicated tracks created by the reconstruction. 
This procedure compares the parameters and correlation matrices of the reconstructed 
tracks and where two are found to be similar, in this case with
a symmetrized KL divergence less than 5000, only the
one with the higher track fit quality is considered.

The inclusive \Bhh sample is the main control sample 
for the determination from data of the probability distribution function (PDF) of the BDT output. 
This sample is selected in exactly the same way as the \Bmm signals
apart from the muon identification requirement.
The same selection is also applied to the \BdKpi normalization channel.

The muon identification efficiency is uniform within $\sim 1\%$ in the considered phase space
therefore no correction is added to the BDT PDF extracted from the \Bhh sample. 
The remaining phase space dependence of the muon identification efficiency 
is instead taken into account in the computation of the normalization factor 
when the \BdKpi channel is considered.

The $J/\psi \to \mu \mu$ decay in the \BuJpsiK and \BsJpsiPhi normalization 
channels is selected in a very similar way to the \Bmm channels, apart from the pointing requirement.
$K^{\pm}$  candidates are required to be identified by the RICH detector and to pass track quality
and impact parameter cuts.

To avoid pathological events, all tracks from selected candidates are required to have a momentum 
less than 1\,TeV/$c$. Only \B candidates with
decay times less than $5 \,\tau_{B_{(s)}^0}$, where $\tau_{B^0_{(s)}}$ is the \B lifetime \cite{PDG}, are accepted for further analysis.
Di-muon candidates coming from elastic di-photon production are  
removed by  requiring  a minimum transverse momentum of the $B$
candidate of 500\mevc.

\section{Determination of the mass and BDT distributions}
\label{sec:bdt}

The variables entering the BDT discriminant are the 
six variables used as input to the \gl
in the 2010 analysis plus three new variables.
The six variables used in the 2010 analysis are the $B$ lifetime, impact parameter, transverse momentum, 
the minimum impact parameter significance (${\rm IP}/\sigma({\rm IP})$) of the muons, the distance of closest approach 
between the two muons and the isolation of the two muons with respect to any other track in the event.
The three new variables are:

\begin{enumerate}
\item {the minimum $p_{\rm T}$ of the two muons;}

\item the cosine of the angle between the muon momentum in the \B rest frame 
and the vector perpendicular to the \B momentum and the beam axis:
\begin{equation}
 \cos P 
= \frac{ p_{y,\mu1} \, p_{x,\B} - p_{x,\mu1} \, p_{y,\B}}{p_{{\rm T},\B} \, (m_{\mu \mu}/2)}
\label{eq:cosy}
\end{equation}
where $\mu_1$ labels one of the muons
and $m_{\mu\mu}$ is the reconstructed \B candidate mass\footnote{As the \B is a (pseudo)-scalar particle, 
this variable is uniformely distributed for signal candidates while is peaked at 
zero for \bbdim background candidates. 
In fact, muons from semi-leptonic decays are mostly emitted in the direction of the $b$'s and,
therefore, lie in a plane formed by the \B momentum and the beam axis.};

\item {the \B isolation} \cite{cdf_iso}
\begin{equation}
 I_{B} = \frac{ p_{\rm T}(B)}{p_{\rm T}(B) + \sum_{i} p_{{\rm T},i}}, 
\label{eq:iso}
\end{equation}
where $p_{\rm T}(B)$ is the \B transverse momentum with respect to the beam line and the sum is over all the tracks,
excluding the muon candidates, that satisfy
$\sqrt{\delta \eta^2 + \delta \phi^2}~<~1.0 $, where $\delta \eta$ and $\delta \phi$ 
denote respectively
the difference in pseudorapidity and azimuthal angle between the track and the \B candidate.

\end{enumerate}

The BDT output is found to be independent of the invariant mass for both 
signal and background and is defined such that the signal is uniformly 
distributed  between zero and one and the background peaks at zero.
The BDT range is then divided in four bins of equal width.
The BDT is trained using simulated samples (\Bmm for signals and \bbdim for 
background where $X$ is any other set of particles)
and the PDF obtained from data as explained below.

\subsection{Combinatorial background PDFs}

The BDT and invariant mass shapes for the combinatorial background inside the signal
regions are determined from data by interpolating the number of  
expected events using  the invariant mass sidebands for each BDT bin.
The boundaries of the signal regions are defined as $m_{\Bd} \pm 60$ \mevcc and $m_{\Bs} \pm 60$ \mevcc and the 
mass sidebands as $[m_{\Bd} - 600 \mevcc, m_{\Bd} - 60 \mevcc]$
and $[m_{\Bs}+60 \mevcc, m_{\Bs}+600 \mevcc]$.

\begin{figure}[t]
\centering
\includegraphics[width=.43\textwidth]{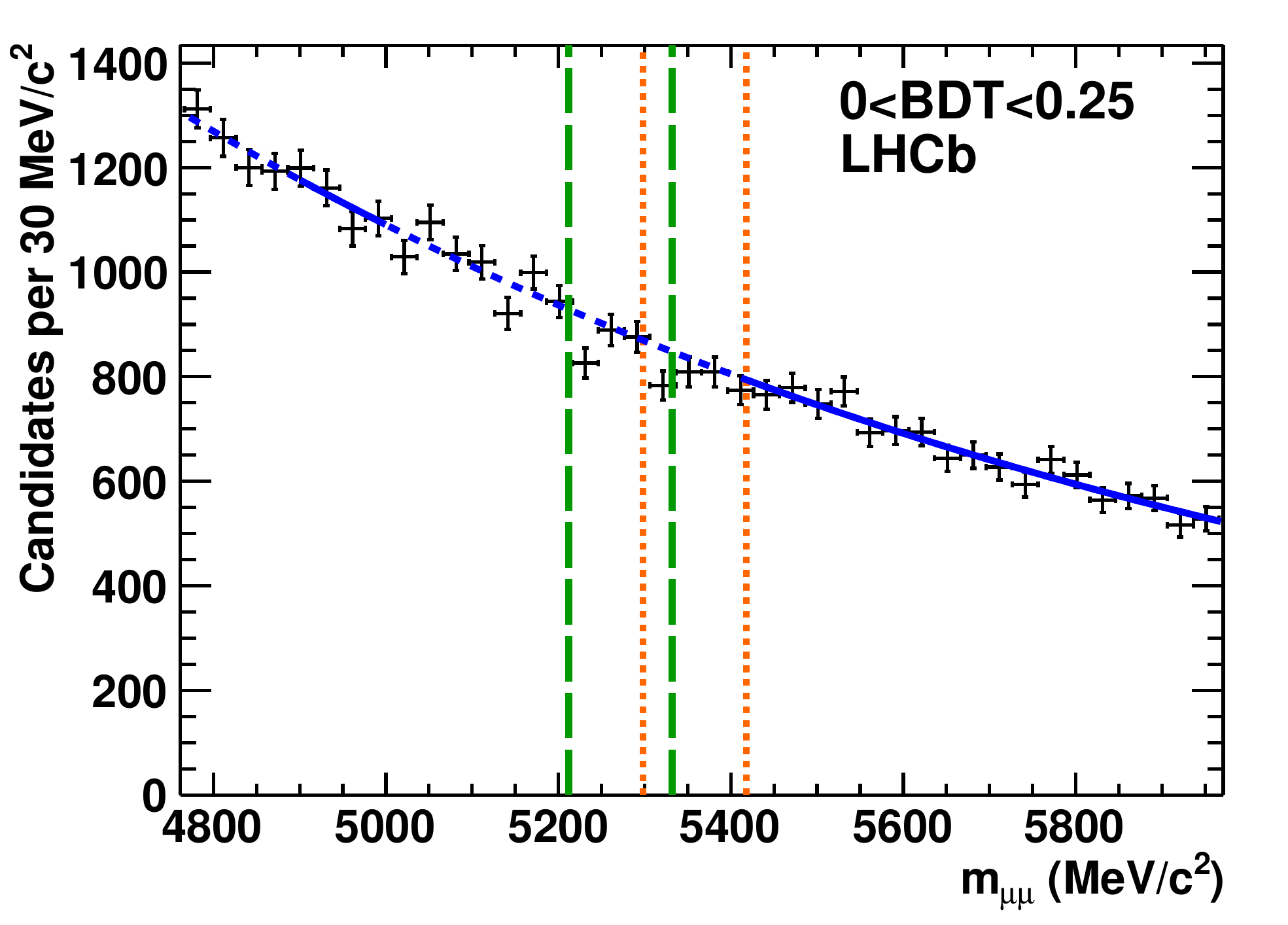}
\includegraphics[width=.43\textwidth]{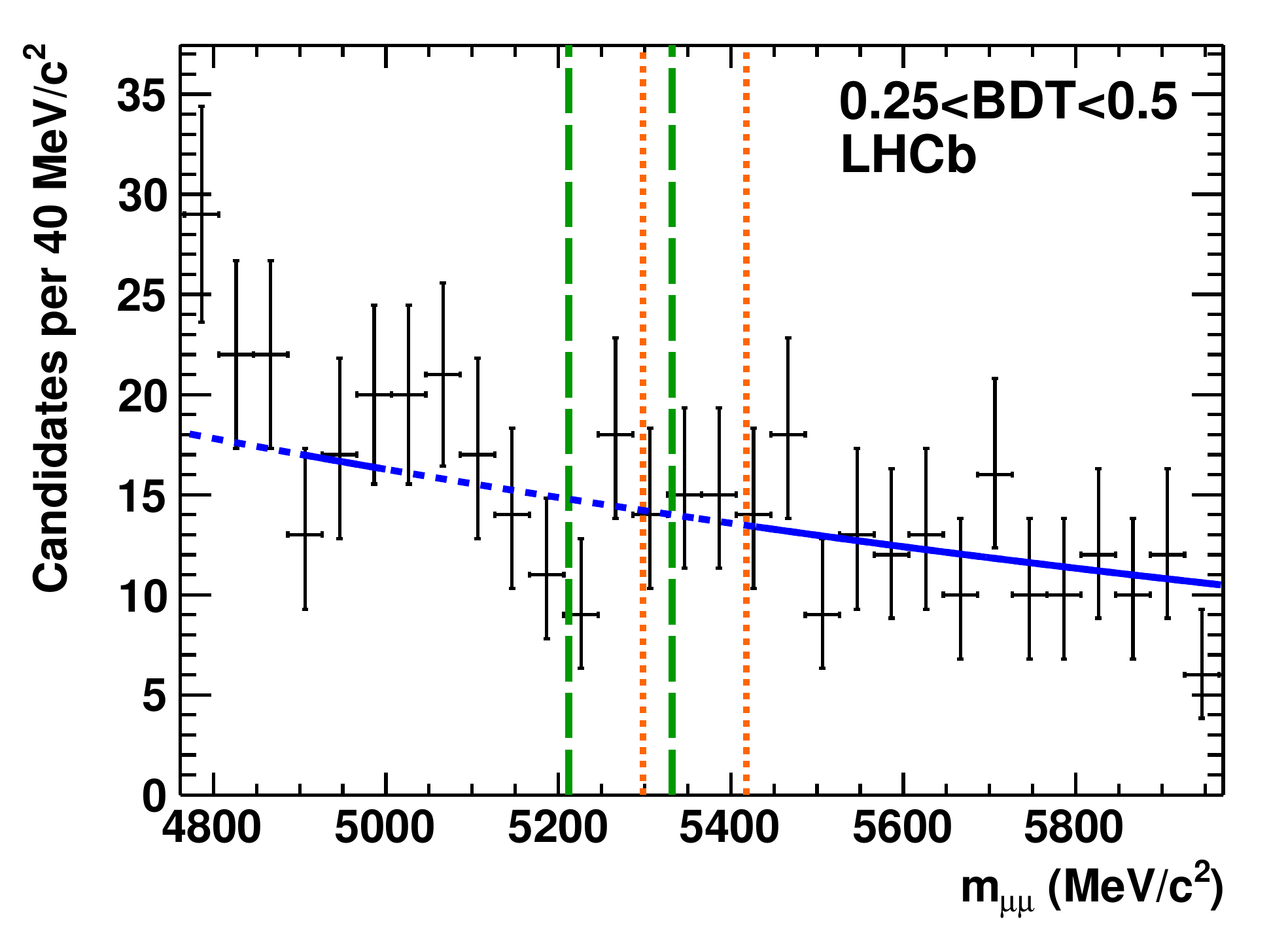}
\includegraphics[width=.43\textwidth]{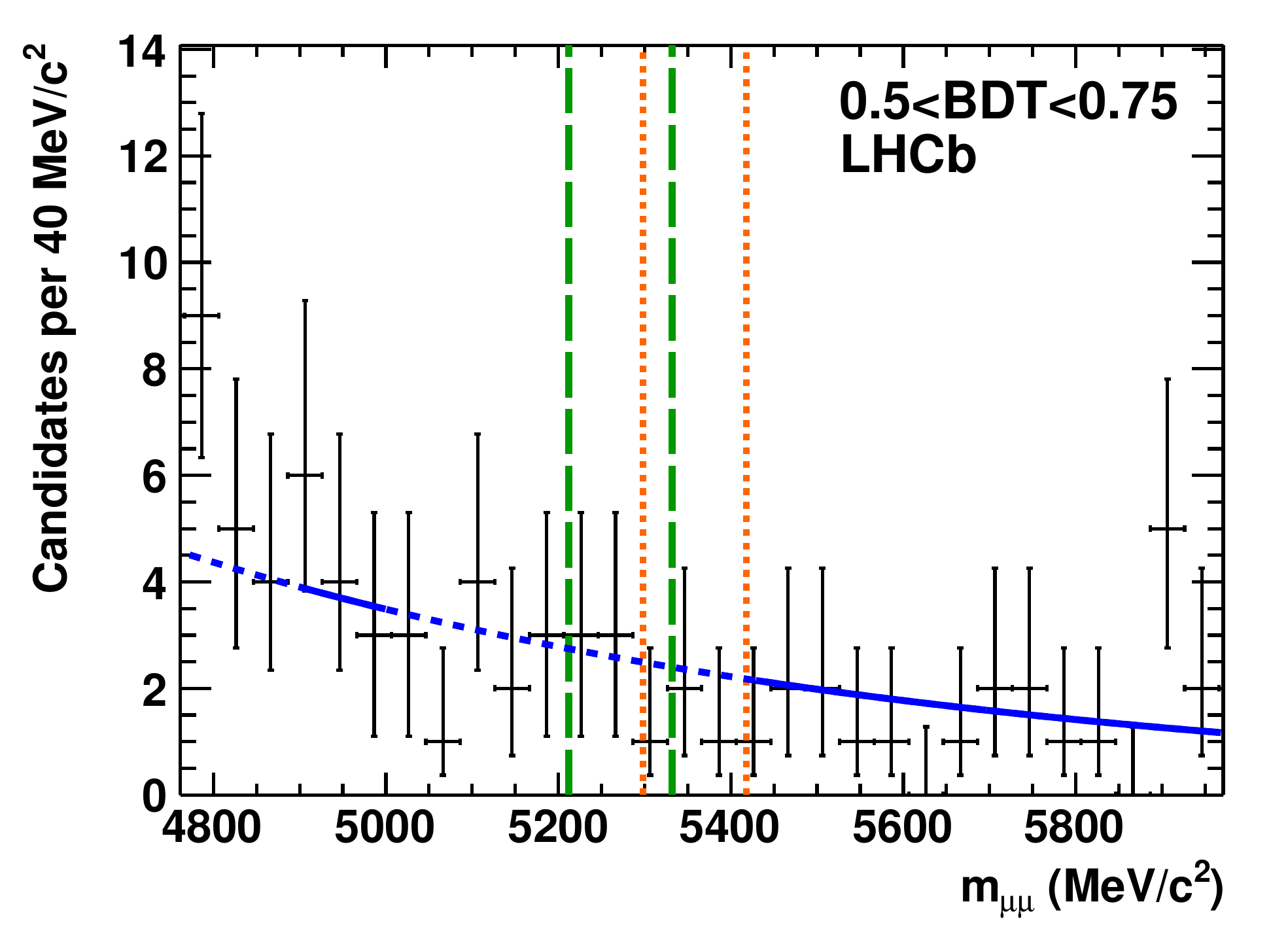}
\includegraphics[width=.43\textwidth]{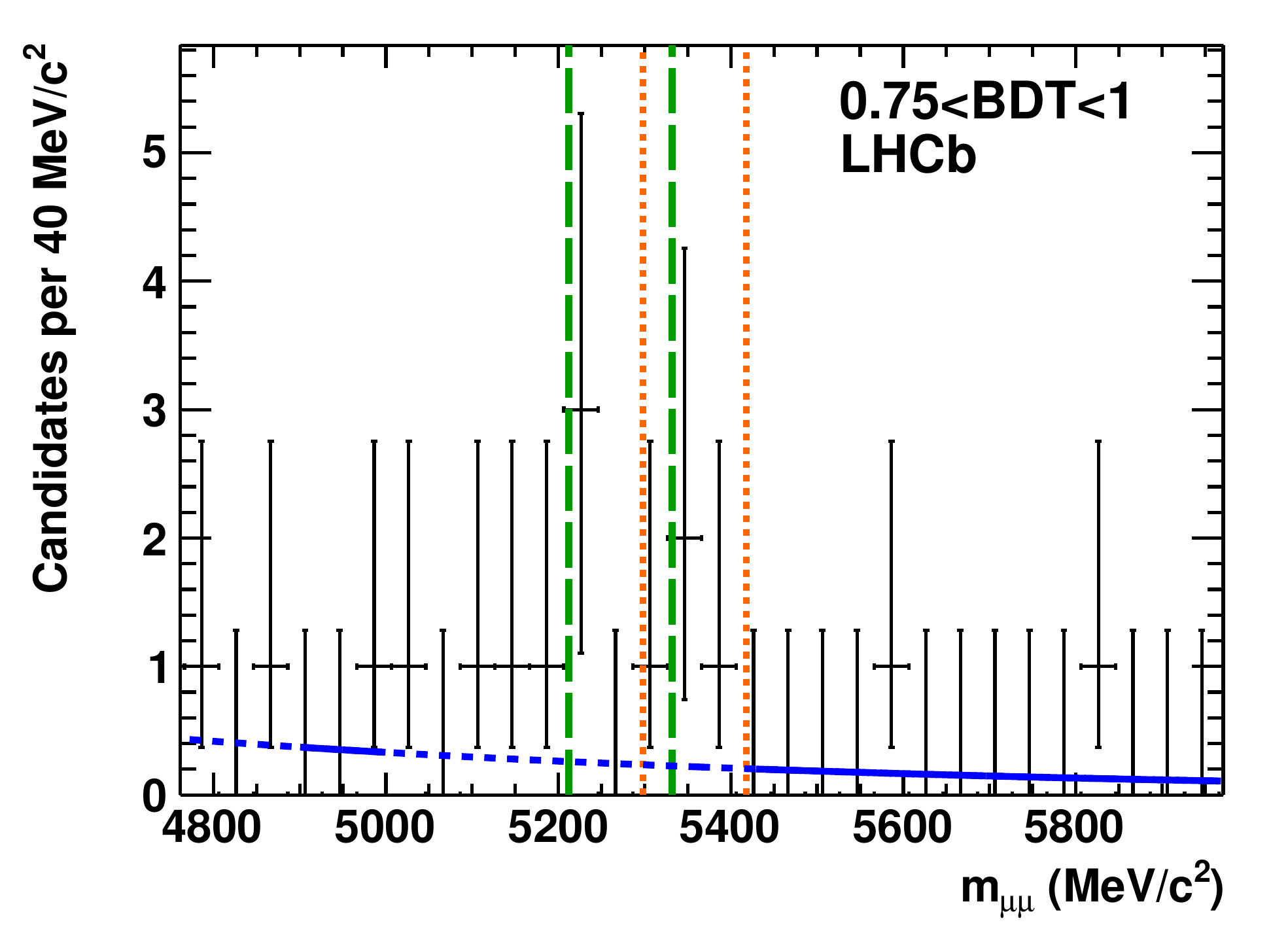}
\vspace{-4mm}
\caption{%
Distribution of the $\mu^+\mu^-$ invariant mass for events in each BDT output bin.
The curve shows the model used to fit the sidebands and extract the expected number 
of combinatorial background events in the \Bs and \Bd signal regions,
delimited by the vertical dotted orange and dashed green lines respectively.
Only events in the region in which the line is solid have been considered in the fit.}
\label{fig:BDT_bkg_bins}
\end{figure}

Figure~\ref{fig:BDT_bkg_bins} shows the invariant mass distribution for events that lie in each BDT output bin. 
In each case the fit model used to estimate the expected number of combinatorial 
background events in the signal regions is superimposed. 

Aside from combinatorial background, the low-mass sideband is potentially polluted by two 
other contributions: cascading $b \to c \mu \nu \to \mu \mu X$ 
decays below 4900\mevcc and peaking background from \Bhh candidates with the two hadrons misidentified 
as muons above 5000\mevcc. To avoid these contaminations,
the number of expected combinatorial 
background events is obtained by fitting a single exponential function
to the events in the reduced low-mass sideband [4900, 5000] \mevcc and in the full high-mass sideband.
As a cross-check, two other models, a single exponential function and 
the sum of two exponential functions, 
have been used to fit the events in different ranges of sidebands providing consistent
background estimates inside the signal regions.

\subsection{Peaking background PDFs}
The peaking backgrounds due to \Bhh events in which both hadrons are 
misidentified as muons
have been evaluated from data and simulated events to be $N_{\Bs} = 1.0\pm 0.4$ 
events and $N_{\Bd} = 5.0 \pm 0.9$ events within the two mass windows and
in the whole BDT output range.
The mass line shape of the peaking background is obtained from a simulated sample of 
doubly-misidentified \Bhh events and normalized to the number of 
events expected in the two search windows from data, $N_{\Bs}$ and $N_{\Bd}$.
The BDT PDF of the peaking background is assumed to be the same as for the signal.

\subsection{Signal PDFs}

The BDT PDF for signal events is determined using an inclusive \Bhh 
sample. Only events which are triggered independently on the signal candidates have been considered (TIS events).

\begin{figure}[t]
\centering
\includegraphics[width=.49\textwidth]{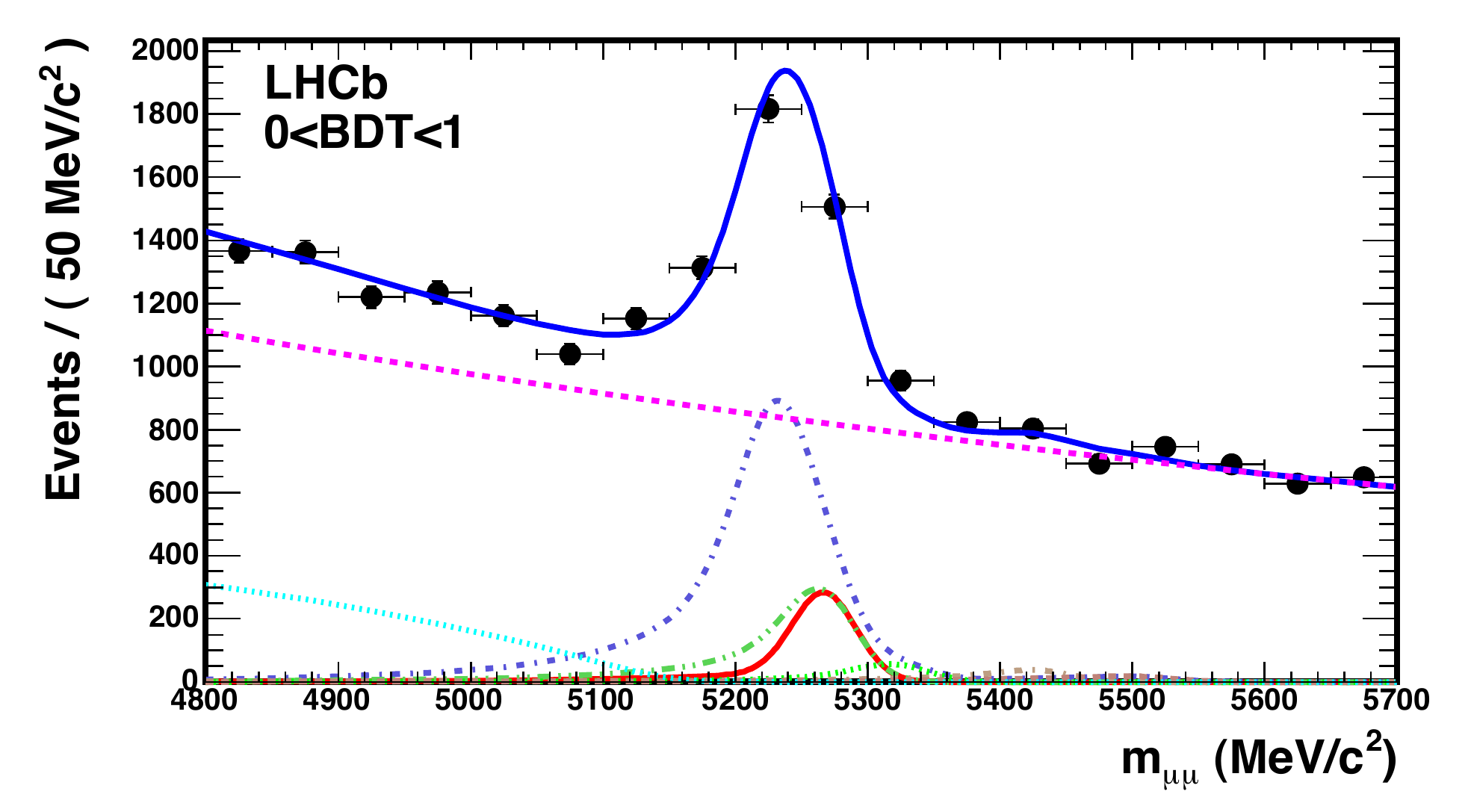}
\includegraphics[width=.49\textwidth]{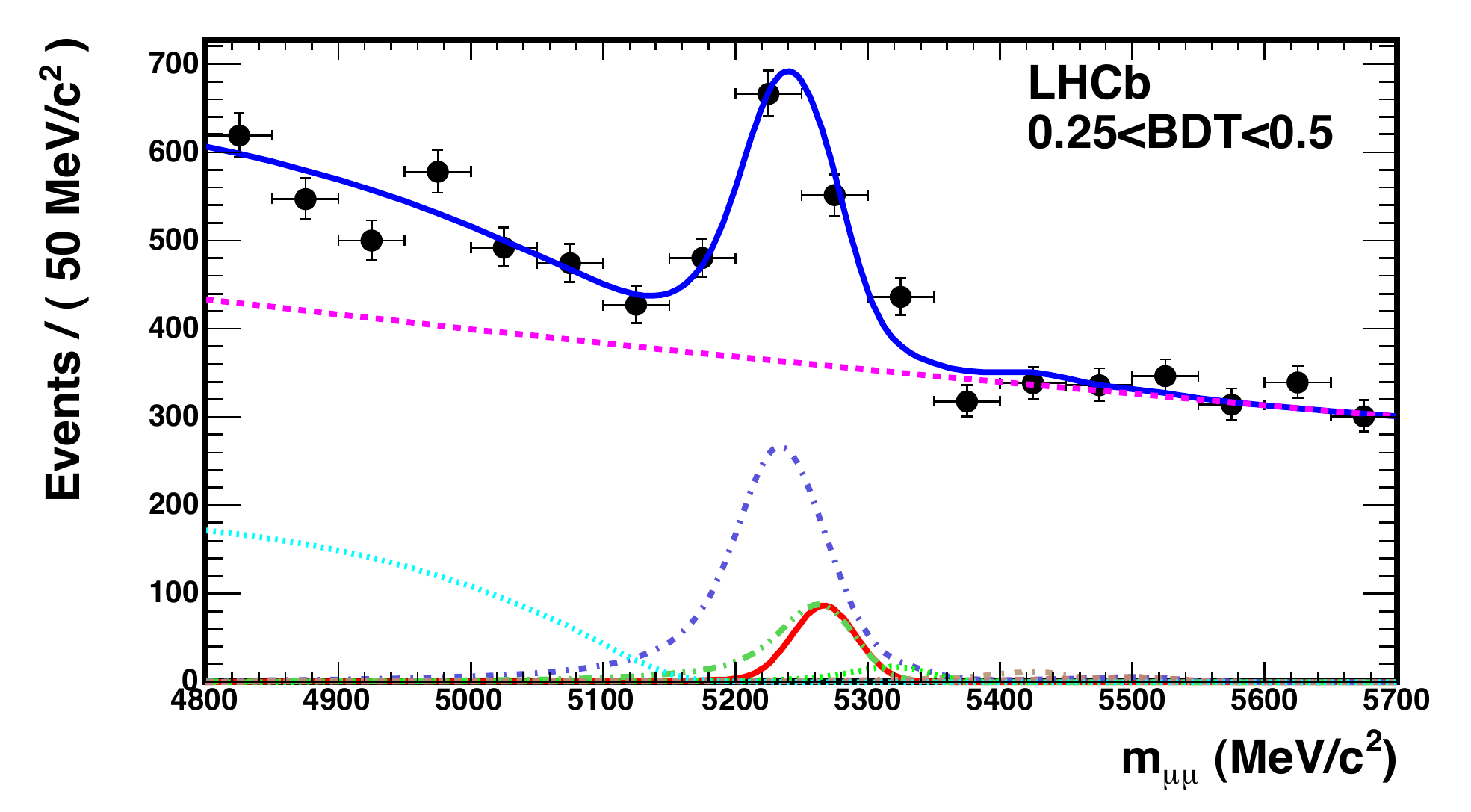}
\includegraphics[width=.49\textwidth]{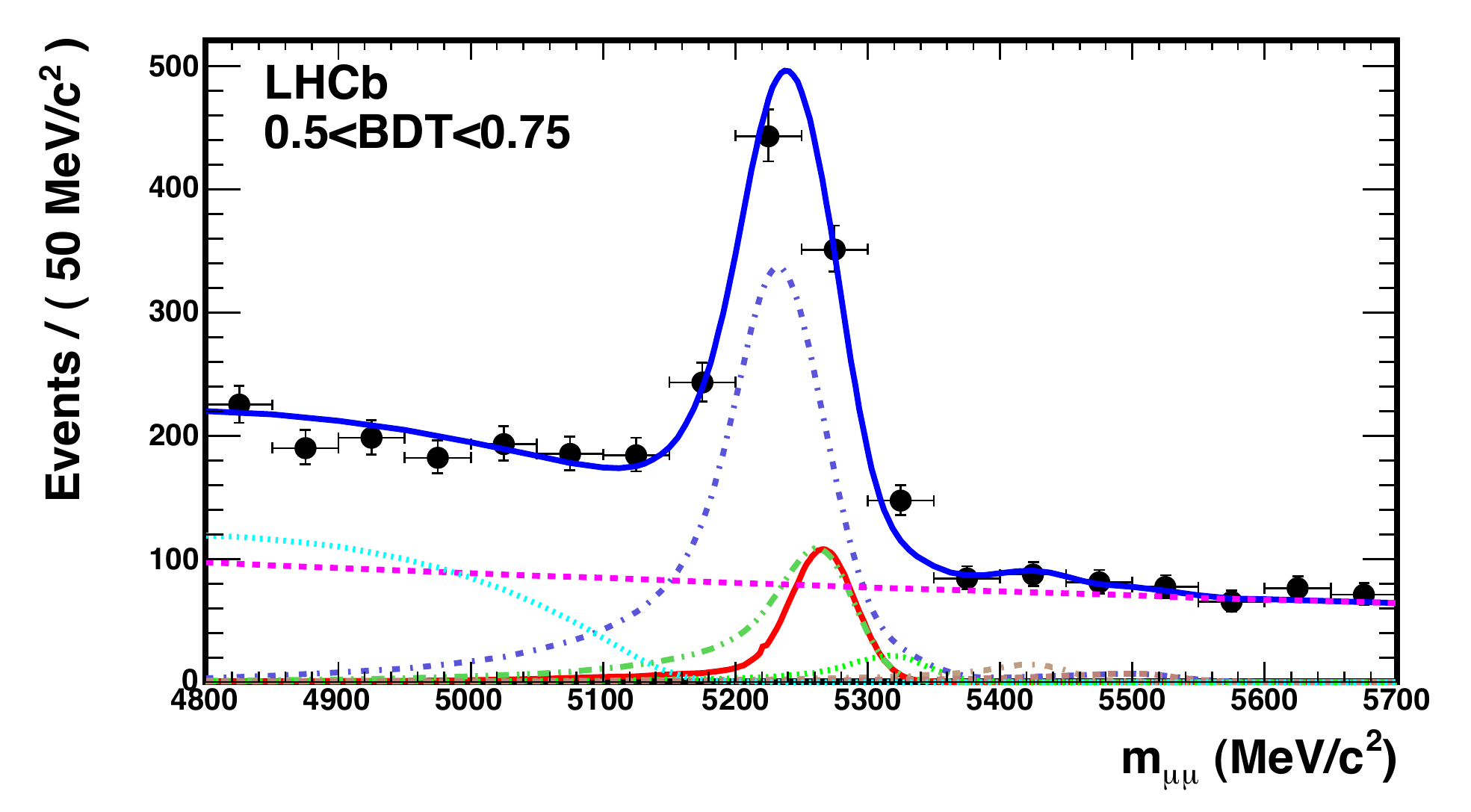}
\includegraphics[width=.49\textwidth]{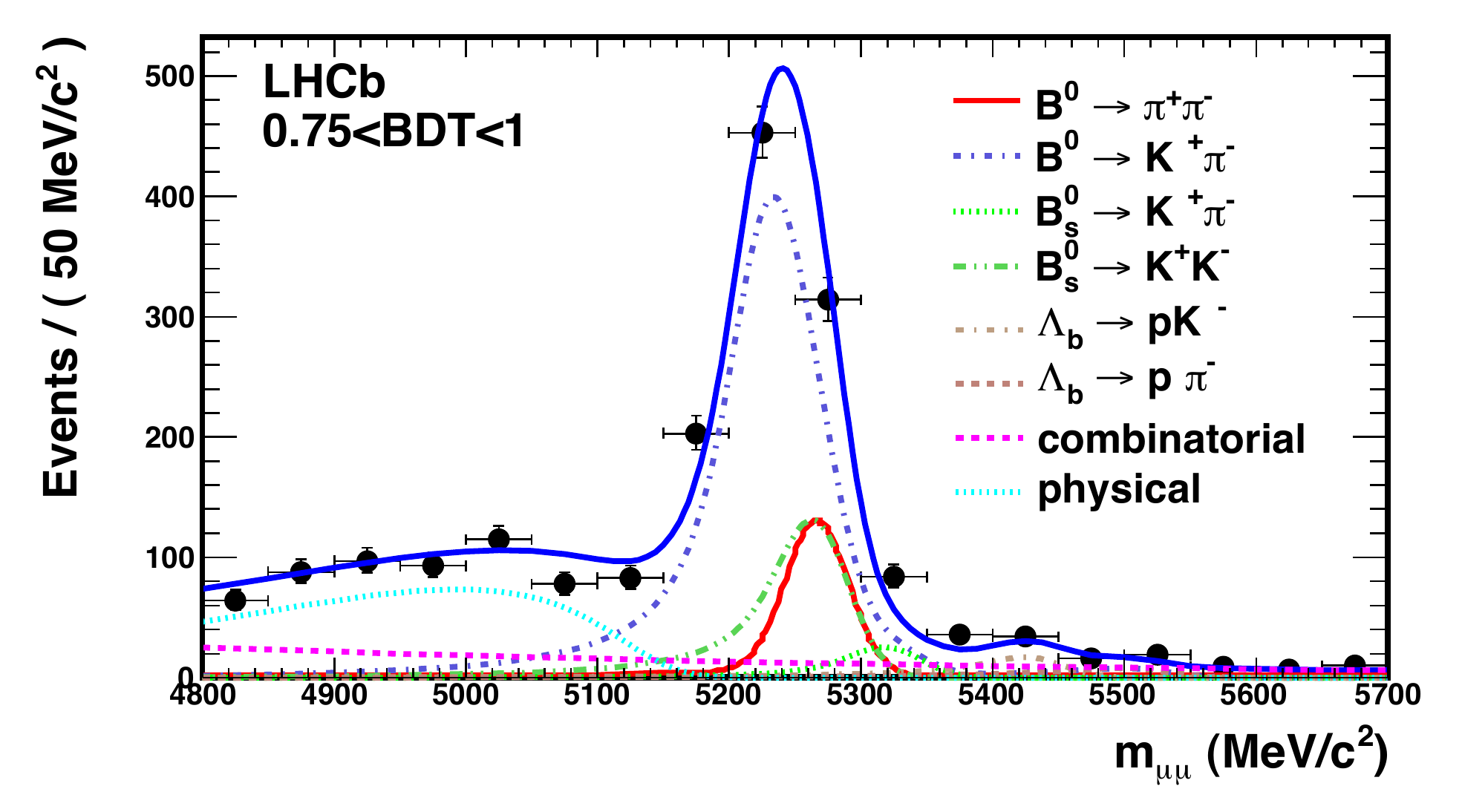}
\vspace{-4mm}
\caption{Invariant mass distributions of \Bhh candidates 
in the $\mu^+\mu^-$ mass hypothesis for the whole sample (top left) and for the samples
in the three highest bins of the BDT output (top right, bottom left, bottom right). 
The \Bhh exclusive decays, the combinatorial background and the
physical background components are drawn under the fit to the data (solid blue line).}
\label{fig:b2hh_fitinbins}
\end{figure}

The number of \Bhh signal events in each BDT output bin is determined by fitting the $hh'$ 
invariant mass distribution under the $\mu\mu$ mass hypothesis \cite{Bhhnote}.
Figure~\ref{fig:b2hh_fitinbins} shows the fit to the mass distribution of the
full sample and for the three highest BDT output bins for \Bhh TIS events.
The \Bhh exclusive decays, the combinatorial background and the
physical background components are drawn under the fit to the data;
the physical background is due to the partial reconstruction of
three-body $B$ meson decays.

\begin{figure}[t]
\centering
\includegraphics[width=.7\textwidth]{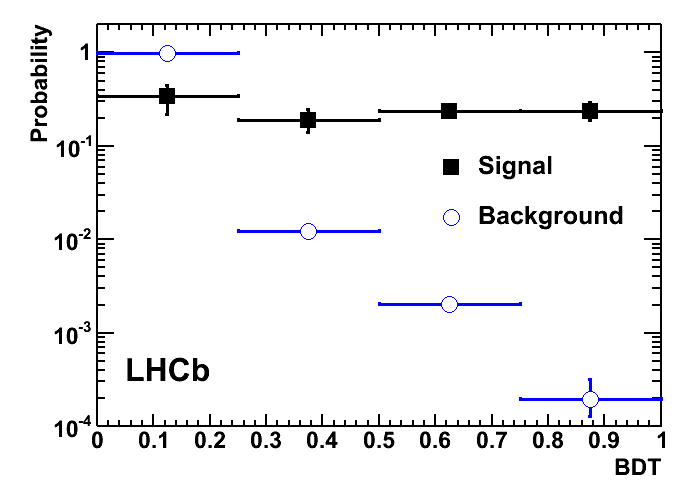}
\vspace{-4mm}
 \caption{BDT probability distribution functions of signal events (solid squares) and combinatorial 
background (open circles):
the PDF for the signal is obtained from the inclusive sample of TIS \Bhh events, the PDF for the combinatorial
background is obtained from the events in the mass sidebands.} 
\label{fig:BDT_all}
\end{figure}

In order to cross-check this result, two other fits have been 
performed on the same dataset.
The signal line shape is parametrized either by a single or a double Crystal Ball function \cite{crystalball}, 
the combinatorial background by an exponential function
and the physical background by an ARGUS function \cite{argus}. 
In addition, exclusive $B^0_{(s)} \to \pi^- K^+, \pi^- \pi^+ , K^-K^+$ channels, 
selected using the $K-\pi$ separation capability of the RICH system, are 
used to cross-check the calibration of the BDT output both using the $\pi^- K^+, \pi^-\pi^+,K^-K^+$ 
inclusive yields without separating \B and \Bs and using the \BdKpi exclusive channel alone.
The maximum spread in the fractional yield obtained
among the different models has been used as a systematic uncertainty in the signal BDT PDF.
The BDT PDFs for signals and combinatorial background 
are shown in Fig.~\ref{fig:BDT_all}.

The invariant mass shape for the signal is parametrized as a Crystal Ball function. 
The mean value is determined
using the \BdKpi and \BsKK exclusive channels and the 
transition point of the radiative tail
is obtained from simulated events \cite{LHCb_paper}. 
The central values are
\begin{eqnarray}
 \mBs & = & 5358.0 \pm 1.0 \mevcc , \nonumber \\ 
 \mBd & = & 5272.0 \pm 1.0 \mevcc. \nonumber
\end{eqnarray}

\begin{figure}[t]
\centering
\includegraphics[width=.49\textwidth]{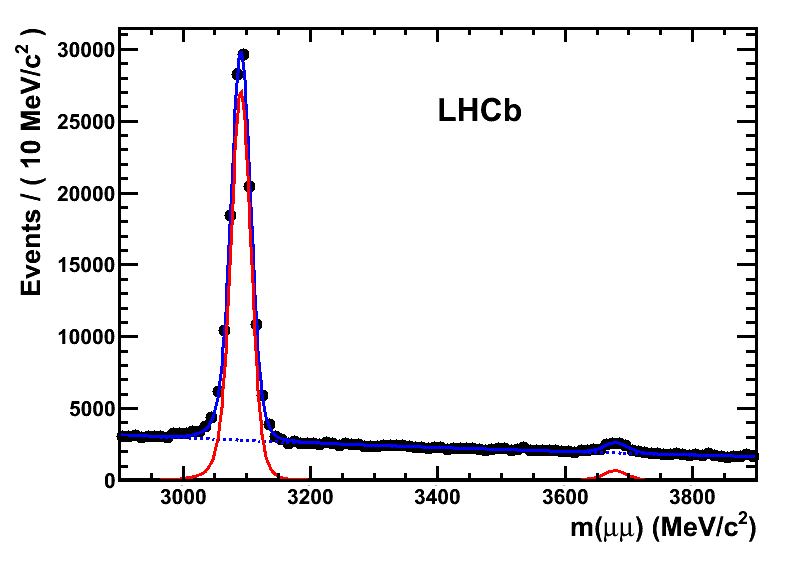}
\includegraphics[width=.49\textwidth]{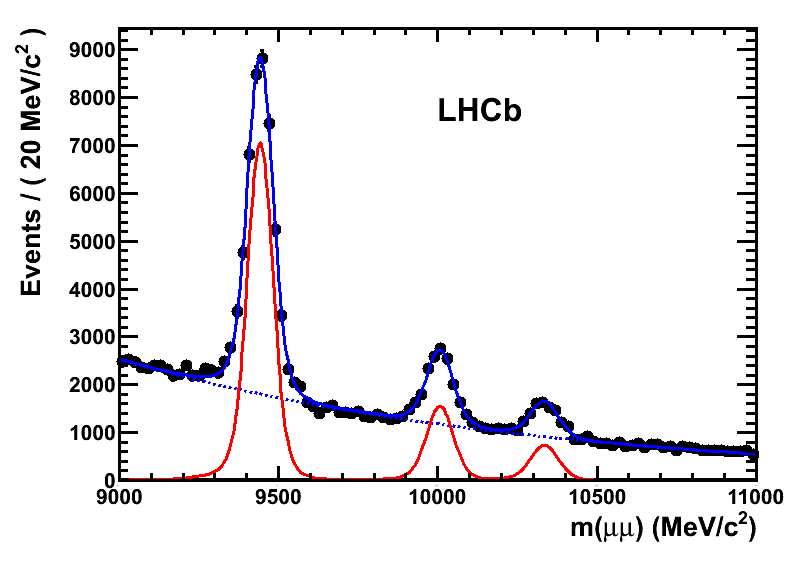}
\vspace{-4mm}
\caption{Di-muon invariant mass spectrum in the ranges (2.9 -- 3.9) \gevcc (left) and \mbox{(9--11) \mevcc} (right).}
\label{fig:dimuon_reso}
\end{figure}

The measured values of \mBd and \mBs are $ 7-8$ \mevcc 
below the PDG values \cite{PDG} due to the fact that the momentum scale is uncalibrated
in the dataset used in this analysis. 
The mass resolutions are extracted from data with  
a linear interpolation between the measured resolution of charmonium and bottomonium 
resonances decaying into two muons: \jpsi, \psitwos, \OneS, \TwoS and \ThreeS.
The mass line shapes for quarkonium resonances are shown in Fig.~\ref{fig:dimuon_reso}.
Each resonance is fitted with two Crystal Ball functions
with common mean value and common resolution but different parameterization of the tails. 
The background is fitted with an exponential function.

The results of the interpolation at the \mBs and \mBd masses are
\begin{eqnarray}
 \sigma(\mBs) & = & 24.6 \pm 0.2_{\rm (stat)} \pm 1.0_{\rm (syst)} \mevcc , \nonumber \\ 
 \sigma(\mBd) & = & 24.3 \pm 0.2_{\rm (stat)} \pm 1.0_{\rm (syst)} \mevcc. \nonumber
\end{eqnarray}

\noindent
This result has been checked using both the fits to the \Bhh inclusive decay 
line shape and the \BdKpi exclusive decay. The results are in agreement within the uncertainties.

\section{Normalization}
\label{sec:efficiencies}

To estimate the signal branching fraction, the number of observed signal events is normalized to 
the number of events of a channel with a well known branching fraction. 
Three complementary normalization channels are used: \BuJpsimumuK, $\Bs \to \jpsi(\mumu)\phi(K^+K^-)$ and 
$\Bd \to K^+ \pi^-$. 
The first two channels have similar trigger and muon identification efficiencies to the signal 
but different number of particles in the final state. The third channel has a similar 
topology but is selected by different trigger lines.

The numbers of \Bsmumu and \Bdmumu candidates are translated 
into a branching fractions (${\cal B}$) using the equation

\begin{equation}
{\cal B} =  {\cal B}_{\rm norm} \times\frac{\rm
\epsilon_{norm}^{REC}
\epsilon_{norm}^{SEL|REC}
\epsilon_{norm}^{TRIG|SEL}
}{\rm
\epsilon_{sig}^{REC}
\epsilon_{sig}^{SEL|REC}
\epsilon_{sig}^{TRIG|SEL}
}\times\frac{f_{\rm norm}}{f_{d(s)}}
\times\frac{N_{\Bmm}}{N_{\rm norm}} 
= \alpha^{\rm norm}_{\Bmm} \times N_{\Bmm},
\label{eq:normalization}
\end{equation}

\noindent where $f_{d(s)}$ and $f_{\rm norm}$ are the probabilities
that a $b$ quark fragments into a $B^0_{(s)}$ and into the $b$ hadron involved
for the chosen normalization mode. 
LHCb has measured $f_s/f_d = 0.267^{+0.021}_{-0.020}$~\cite{fdfs}. 
${\cal B}_{\rm norm}$ is the branching fraction 
and $N_{\rm norm}$ is the number of selected events of the normalization channel. 
The efficiency is the product of three factors: $\epsilon^{\rm REC}$ is the reconstruction efficiency of 
all the final state particles of the decay including the geometric acceptance of the
detector; $\epsilon^{\rm SEL|REC}$ is the selection efficiency for reconstructed events; 
$\epsilon^{\rm TRIG|SEL}$ is the trigger efficiency for
reconstructed and selected events. The subscript ($\rm sig,norm$) indicates whether the efficiency 
refers to the signal or the normalization channel. Finally, $\alpha^{\rm norm}_{\Bmm}$ is the normalization 
factor (or single event sensitivity) and $N_{\Bmm}$ the number of observed signal events.

\begin{figure}[t] 
\begin{center}
  \includegraphics*[width=0.49\textwidth]{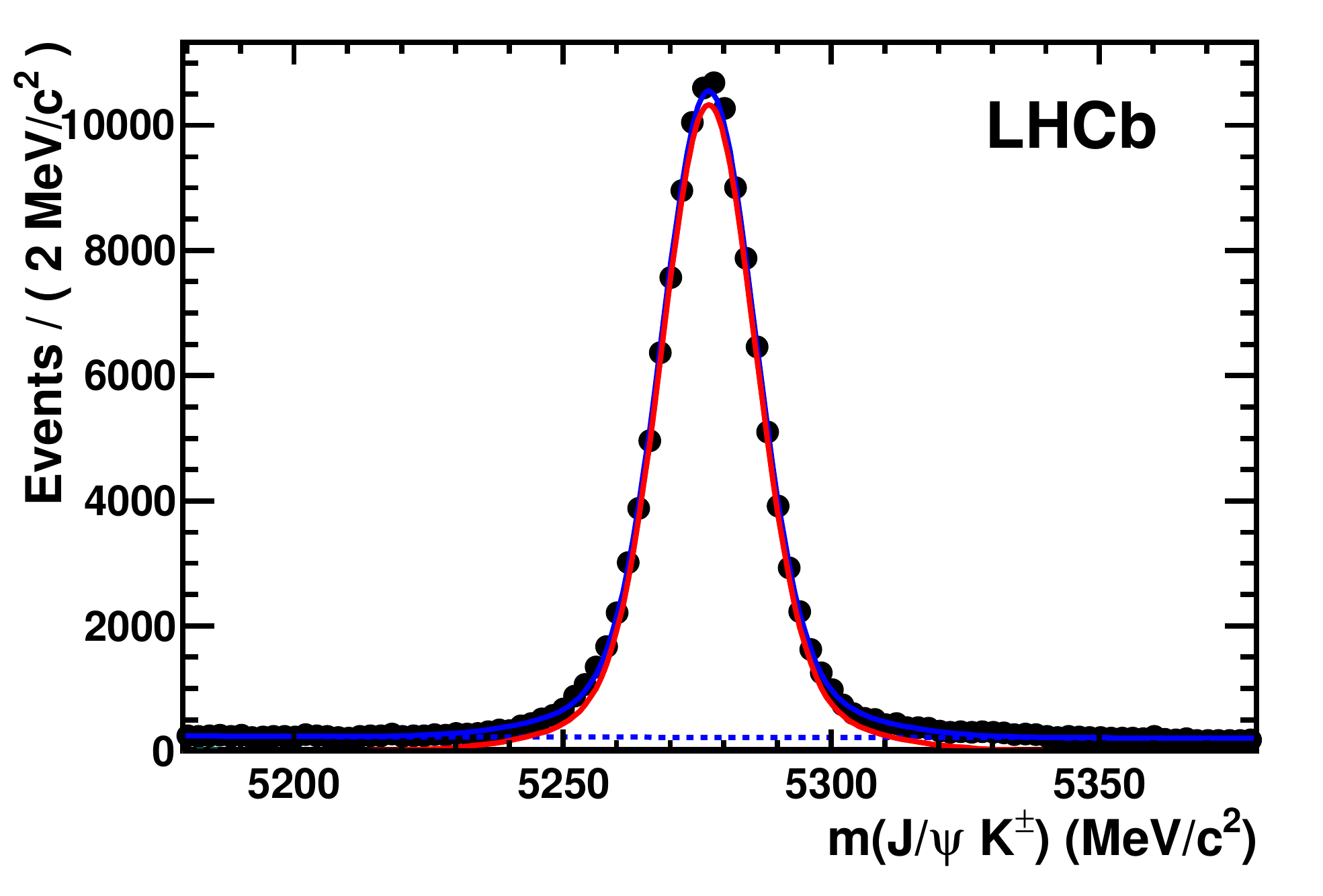}
  \includegraphics*[width=0.49\textwidth]{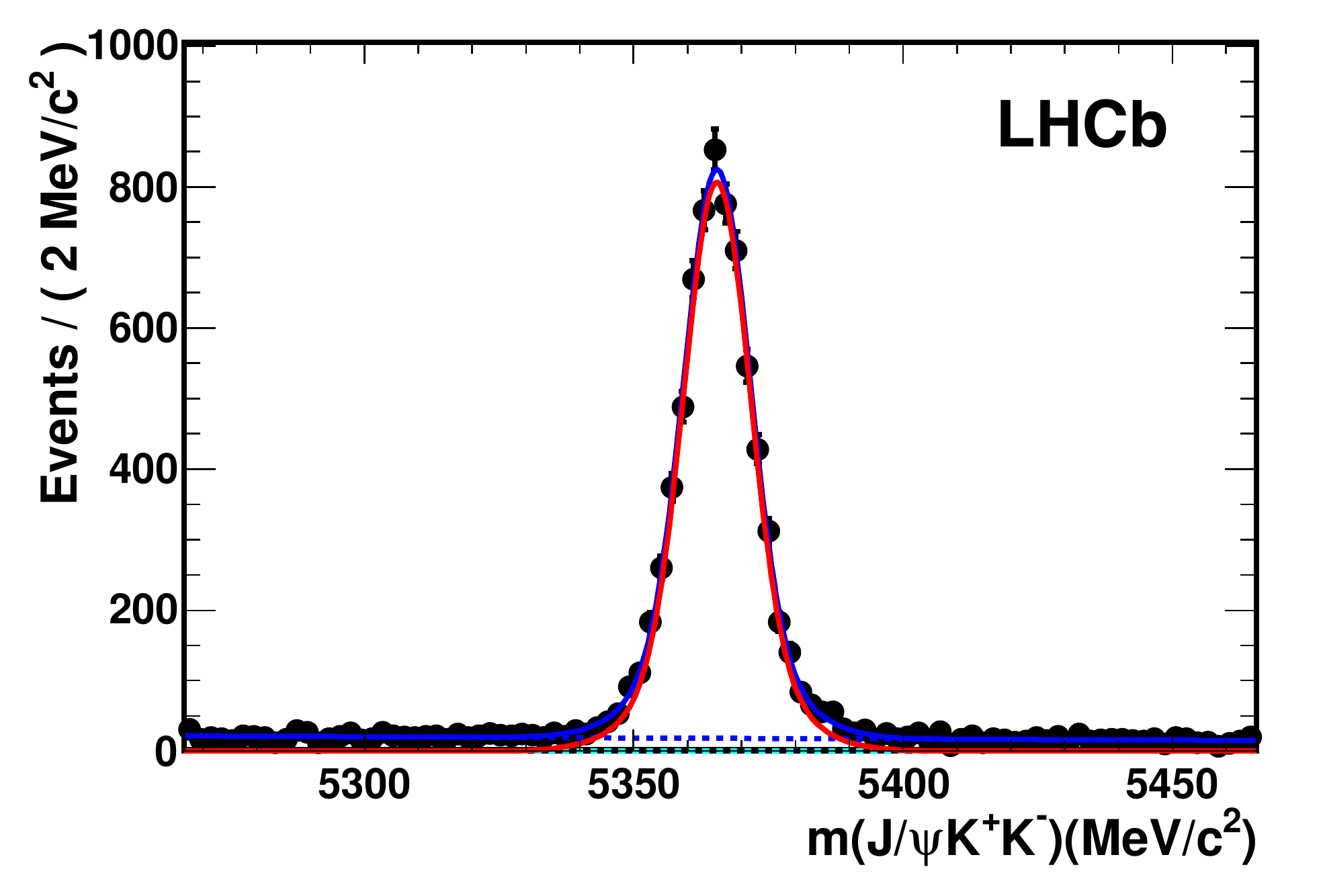}
\vspace{-4mm}
  \caption
  {Invariant mass distributions of the \BuJpsiK (left) and \BsJpsiPhi (right) candidates used in the normalization procedure.}  
  \label{fig:num_Bu}
\end{center}
\end{figure}

For each normalization channel $N_{\rm norm}$ is obtained from a fit to the 
invariant mass distribution.
The invariant mass distributions for reconstructed \BuJpsiK and \BsJpsiPhi candidates 
are shown in Fig.~\ref{fig:num_Bu}, while the $B^0 \to K^+ \pi^-$ yield is obtained 
from the full \Bhh fit as shown in the top left of Fig.~\ref{fig:b2hh_fitinbins}.

The numbers used to calculate the normalization factors are 
summarized in Table~\ref{tab:norm_summary}.
A weighted average of the three normalization channels, 
assuming the tracking and trigger uncertainties to be correlated between the 
two \Jpsi normalization channels and the uncertainty on $f_d/f_s$ 
to be correlated between the \BuJpsiK and \BdKpi, 
gives
\begin{eqnarray}
\alpha^{\rm norm}_{\Bsmumu}= (8.38 \pm 0.74) \times 10^{-10}\, , \nonumber \\ 
\alpha^{\rm norm}_{\Bdmumu}= (2.20 \pm 0.11) \times 10^{-10}\, . \nonumber 
\end{eqnarray} 
These normalization factors are used to determine the limits.

\begin{table}
\begin{center}
\tabcaption{Summary of the quantities and their uncertainties required to calculate the normalization factors 
($\alpha^{\rm norm}_{\Bmm}$) 
for the three normalization channels considered. 
The branching fractions are taken from Refs.~\cite{PDG,BELLE_Bs}. 
The trigger efficiency and the number of \BdKpi candidates correspond to TIS events.}
\vspace{3mm}
\footnotesize
    \label{tab:norm_summary}
    \begin{tabular}{cr@{}lr@{}lr@{}lr@{}lr@{}lr@{}l}
\toprule
                   &   \multicolumn{2}{c}{\BR}     &   
\multicolumn{2}{c}{$\frac{\rm
\epsilon_{norm}^{REC}
\epsilon_{norm}^{SEL|REC}
}{\rm
\epsilon_{sig}^{REC}
\epsilon_{sig}^{SEL|REC}
}$}      &   
\multicolumn{2}{c}{$\frac{\rm
\epsilon_{norm}^{TRIG|SEL}\TTstrut
}{\rm
\epsilon_{sig}^{TRIG|SEL} \BBstrut
}$}  &   
 
\multicolumn{2}{c}{$N_{\rm norm}$} &  
\multicolumn{2}{c}{$\alpha^{\rm norm}_{B^0 \rightarrow \mu^+ \mu^-}$} & 
\multicolumn{2}{c}{$\alpha^{\rm norm}_{B^0_{s} \rightarrow \mu^+ \mu^-}$} \\
  & \multicolumn{2}{c}{$(\times 10^{-5})$}  &  & & & &  &  &  \multicolumn{2}{c}{$(\times 10^{-10})$} & \multicolumn{2}{c}{$(\times 10^{-9})$}  \\
      \midrule  \midrule
\BuJpsiK \BBstrut\TTstrut& $6.01\,\pm$ & $\,0.21$ & $0.48\,\pm$ & $\,0.014 $ & $0.95\,\pm$ & $\,0.01 $ & 
$124\,518\,\pm$ & $\,2\,025$ & $ 2.23\,\pm$ & $\,0.11$ & $0.83\,\pm$ &$\,0.08 $ \\
\BsJpsiPhi \BBstrut& $3.4\,\pm$ & $\, 0.9$ & $0.24\,\pm$ & $\,0.014 $ & $0.95\,\pm$ & $\,0.01 $ & $6\,940\,\pm$ & $\,93$ &  $2.96\,\pm$ & $\,0.84$ & $1.11\,\pm$ & $\,0.30$ \\
\BdKpi \BBstrut& $1.94\,\pm$ & $\,0.06$ & $0.86\,\pm$ & $\,0.02 $ & $0.049\,\pm$ & $\,0.004$ & 
$4\,146\,\pm$ & $\,608$  &$1.98\,\pm$ & $\,0.34$ & $0.74\,\pm$ & $\,0.14$ \\

\bottomrule
    \end{tabular}
  \end{center}
\normalsize 
\end{table}

\section{Results}
\label{sec:exp_limit}

The results for \Bsmumu and \Bdmumu are summarized in Table~\ref{tab:data_bsmm} and 
Table~\ref{tab:data_bdmm} respectively and in each of the bins the expected number of 
combinatorial background, peaking background, signal events, with the SM prediction assumed, 
is shown together with the observations on the data.
The uncertainties in the signal and background PDFs and normalization factors 
are used to compute the uncertainties on the background 
and signal predictions.

\begin{figure}[tbp]
  \begin{center}
     \includegraphics*[width=0.7\textwidth]{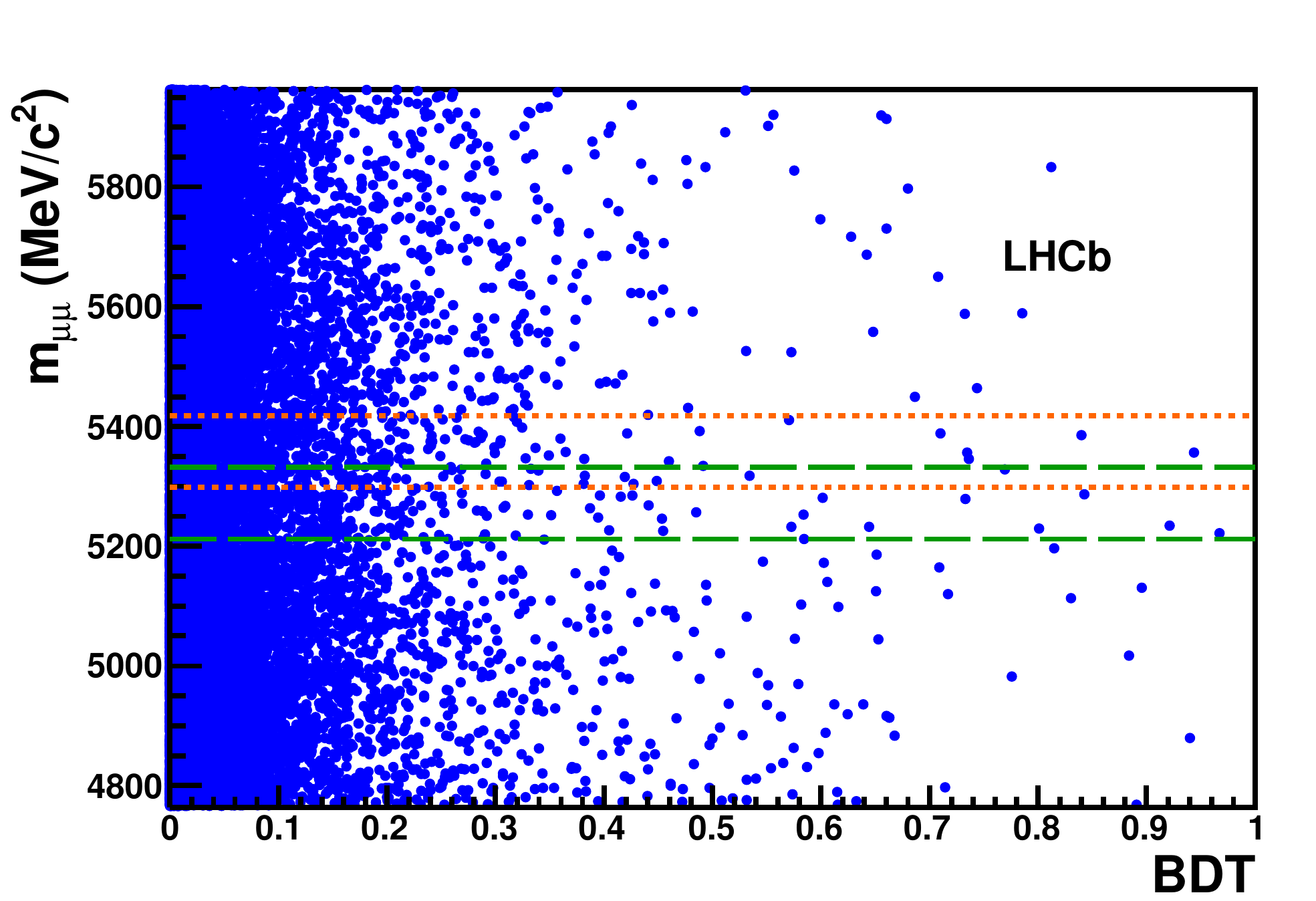}
  \end{center}
\vspace{-4mm}
  \caption{Distribution of selected di-muon events in the 
     invariant mass--BDT plane. The orange short-dashed 
(green long-dashed) lines indicate the $\pm 60 \mevcc$ search window around
    the mean \Bs (\Bd) mass.}
\label{fig:BDTvsmass_obs}
\end{figure}

\begin{figure}[tbp]
  \begin{center}
    \includegraphics*[width=0.9\textwidth]{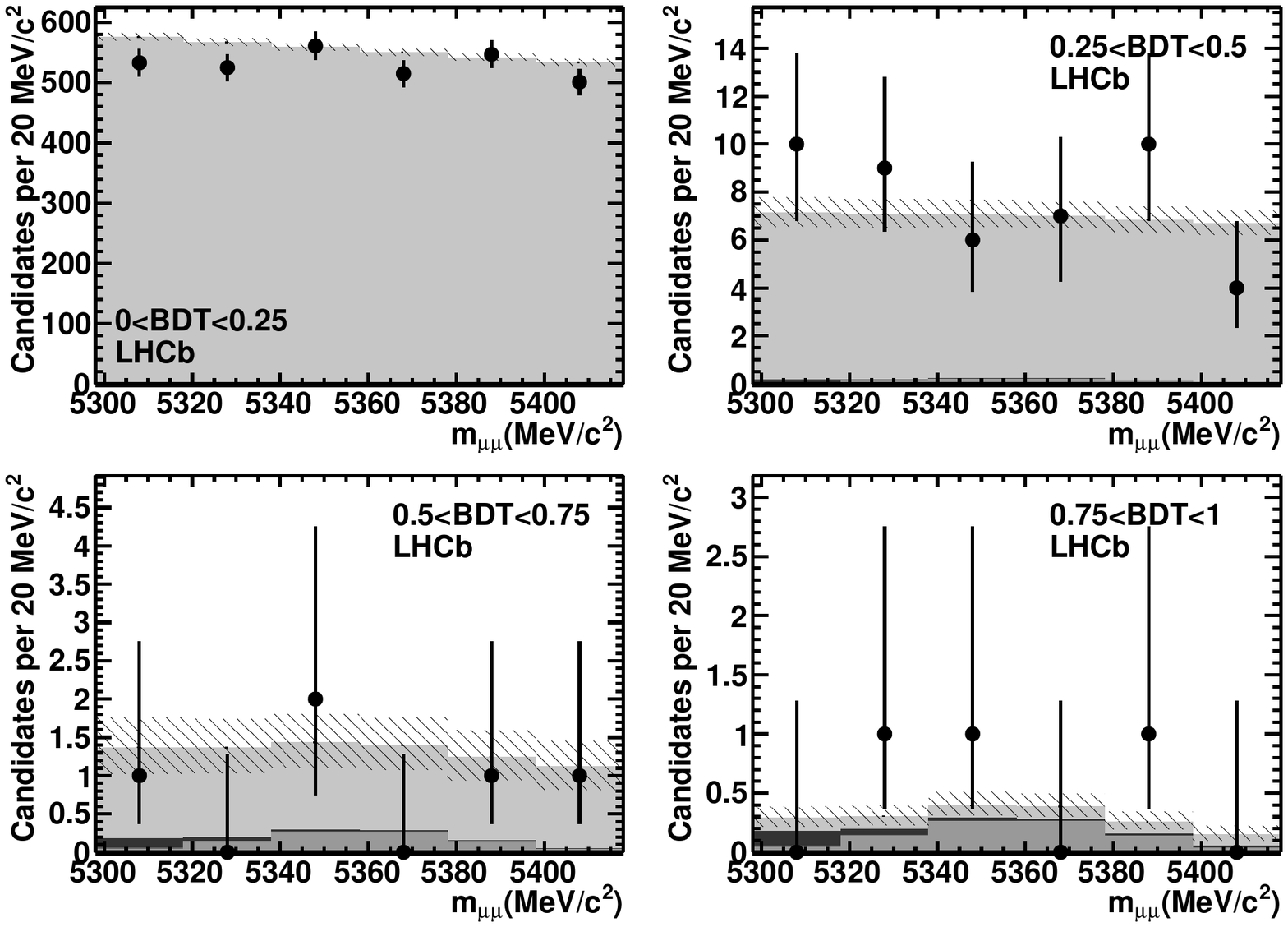}
  \end{center}
\vspace{-4mm}
  \caption{Distribution of selected di-muon events in the \Bsmumu
     mass window for the four BDT output bins. The black dots are data, the light grey histogram shows 
the contribution of the combinatorial background, the black filled histogram shows
the contribution of the \Bhh background and the dark grey filled histogram the contribution of \Bsmumu signal events according to the SM rate. The hatched area depicts the uncertainty on the sum of the expected contributions.} 
\label{fig:fondo_bs}
\end{figure}

\begin{figure}[tbp]
  \begin{center}
    \includegraphics*[width=0.9\textwidth]{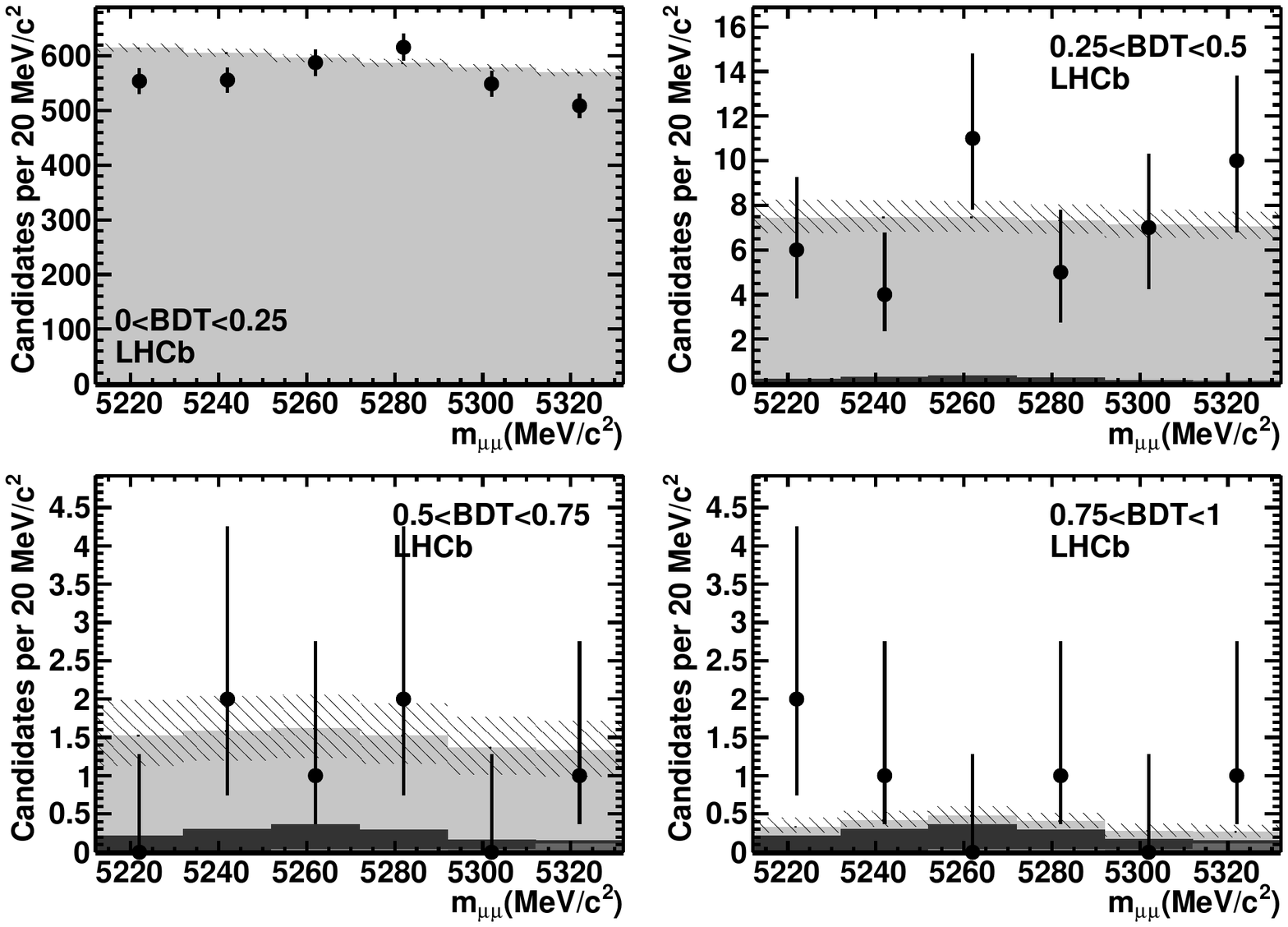}
  \end{center}
\vspace{-4mm}
  \caption{Distribution of selected di-muon events in the \Bdmumu 
     mass window for the four BDT output bins. The black dots are data, the light grey histogram shows 
the contribution of the combinatorial background, the black filled histogram shows
the contribution of the \Bhh background and the dark grey filled histogram shows 
the cross-feed of \Bsmumu events in the \Bd mass window assuming the the SM rate.  The hatched area depicts the uncertainty on the sum of the expected contributions.} 
\label{fig:fondo_bd}
\end{figure}

The two dimensional (mass, BDT) distribution of selected events 
can be seen in Fig.~\ref{fig:BDTvsmass_obs}.
The distribution of the invariant mass in the four BDT bins 
is shown in Fig.~\ref{fig:fondo_bs} for \Bsmumu 
and in Fig.~\ref{fig:fondo_bd} for \Bdmumu selected candidates. 

The compatibility of the distribution of events inside the search window 
in the invariant mass--BDT plane 
with a given branching fraction hypothesis is evaluated using the \CLs method~\cite{Read_02}. 
This method provides three estimators: \CLsb, a measure of the compatibility 
of the observed distribution with the signal and background hypotheses, \CLb, a measure of the compatibility with the 
background-only hypothesis and \CLs, a measure of the compatibility 
of the observed distribution with the signal and background hypotheses 
normalized to the background-only hypothesis.

\begin{figure}[tbp]
\centering
\includegraphics[width=0.49\textwidth, height=5cm]{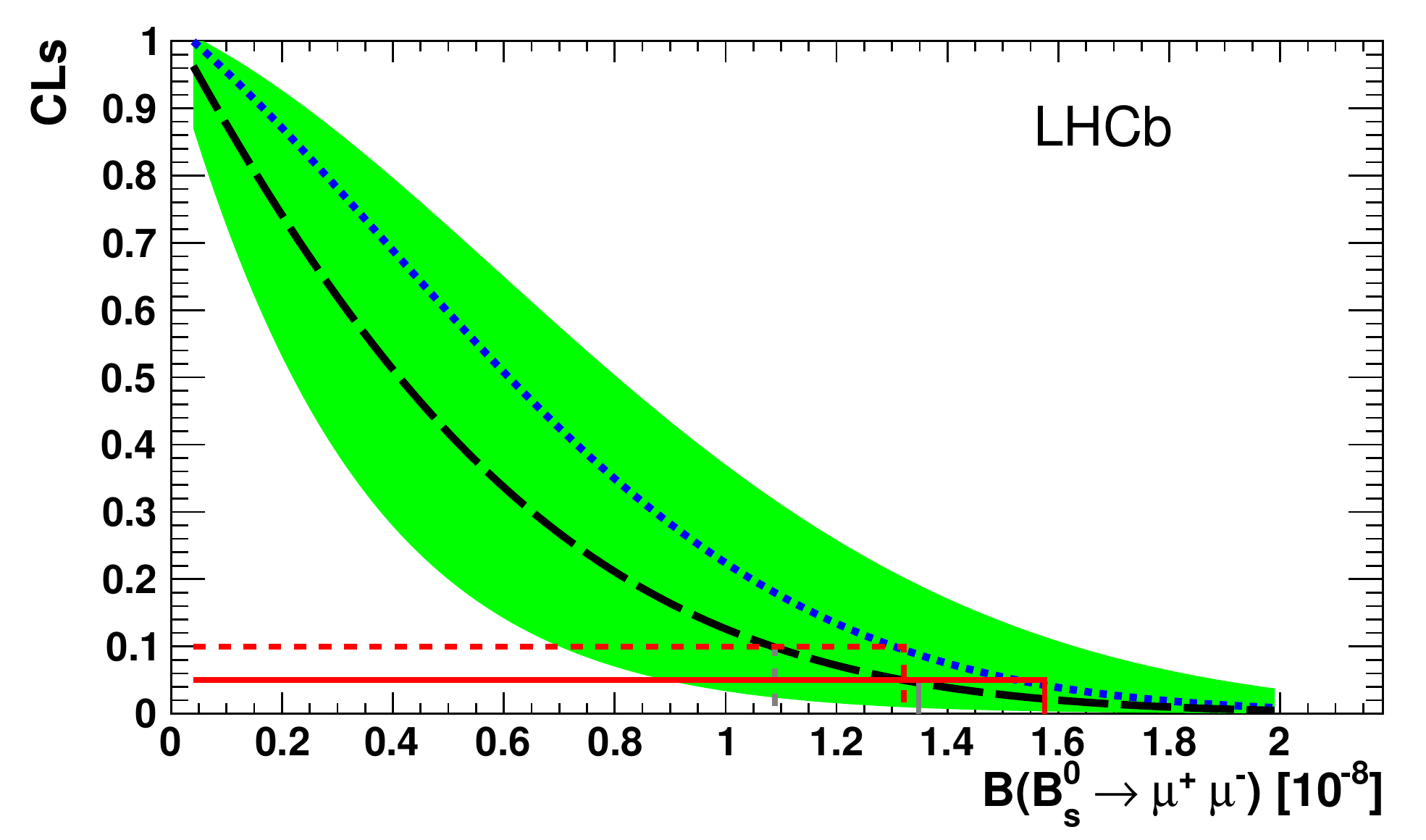}  
\includegraphics[width=0.49\textwidth, height=5cm]{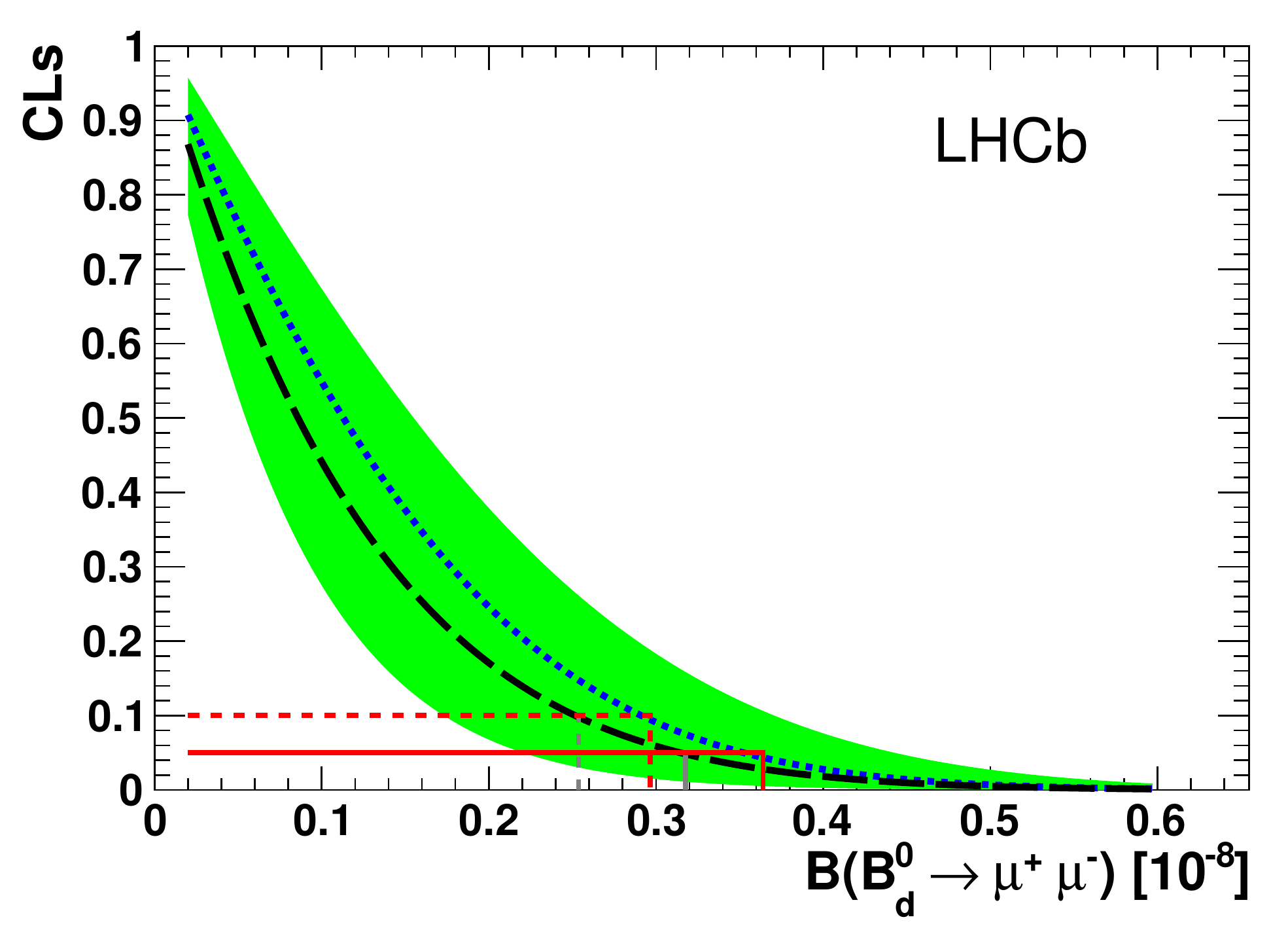}
\vspace{-4mm}
\caption
{ \CLs as a function of the assumed \BR. Expected (observed) values are 
shown by dashed black (dotted blue) lines. The expected \CLs values have been computed 
assuming a signal yield corresponding to the SM branching fractions. 
The green (grey) shaded areas cover the region of $\pm 1 \sigma$ of compatible observations.
The measured upper limits at 90\% and 95\% CL\ are also shown.
Left: \Bsmumu, right: \Bdmumu.}
\label{fig:cls_bsbd}
\end{figure}

The expected \CLs values are shown 
in Fig.~\ref{fig:cls_bsbd} for \Bsmumu and
for \Bdmumu as dashed black lines 
under the hypothesis that background and SM events are observed.
The shaded areas cover the region of $\pm 1 \sigma$ of compatible observations.
The observed values of  \CLs as a function of the assumed branching ratio is shown as dotted blue 
lines on both plots.

The expected limits and the measured limits for \Bsmumu and \Bdmumu at 90\,\% and 95\,\% CL 
are shown in Table~\ref{tab:bs_results} 
and Table~\ref{tab:bd_results}, respectively.
For the \Bsmumu decay, 
the expected limits are computed allowing the 
presence of \Bsmumu events according to the SM branching 
fraction. For the \Bdmumu decay
the expected limit is computed in the background-only hypothesis and also allowing the presence
of \Bdmumu events with the SM rate: the two results are identical.
In the determination of the limits, the cross-feed of \Bsmumu (\Bdmumu) events in the
\Bd (\Bs) mass window has been taken into account assuming the SM rates.

The observed \CLb values are shown in the same tables. 
The comparison of the observed distribution of events with the expected background distribution 
results in a  p-value $(1-\CLb)$ of 5\,\% for the \Bsmumu and 32\,\% for the \Bdmumu decay.
For the \Bsmumu decay, the probability that the observed events are compatible with the 
sum of expected background events 
and signal events according to the SM rate is measured by $1-$\CLsb and it is 33\%.

The result obtained in 2011 with  0.37 fb$^{-1}$ has been combined 
with the published  result based on $\sim 37$ pb$^{-1}$\cite{LHCb_paper}.
The expected and observed limits for 90 \% and 95 \% \CL for the combined results are shown in 
Table~\ref{tab:bs_results} for the \Bsmumu decay and in Table~\ref{tab:bd_results} for the \Bdmm decay.

\begin{table}[t]
\tabcaption{Expected combinatorial background events, expected peaking (\Bhh) background events, 
expected signal events assuming the SM \BF prediction, 
and observed events in the \Bsmumu search window.}
\begin{center}
\resizebox {\textwidth }{!}{%
\begin{tabular}{|l c c|c|c|c|c|}
\hline
& & & \multicolumn{4}{c|}{BDT\TTstrut\BBstrut}  \\
& & & 0. -- 0.25  & 0.25 -- 0.5 & 0.5 -- 0.75 & 0.75 -- 1. \\
\hline
\multirow{30}{*}{\rotatebox{90}{Invariant mass [\MeVcc]}}
         & \multirow{4}{*}{5298  -- 5318 }
                 & Expected comb. bkg & $575.5^{+6.5}_{-6.0}$ & $6.96^{+0.63}_{-0.57}$ & $1.19^{+0.39}_{-0.35}$ & $0.111^{+0.083}_{-0.066}$ \TTstrut\\
                 && Expected peak. bkg & $0.126^{+0.037}_{-0.030}$ & $0.124^{+0.037}_{-0.030}$ & $0.124^{+0.037}_{-0.030}$ & $0.127^{+0.038}_{-0.031}$ \TTstrut\\
                 && Expected signal & $0.059^{+0.023}_{-0.022}$ & $0.0329^{+0.0128}_{-0.0095}$ & $0.0415^{+0.0120}_{-0.0085}$ & $0.0411^{+0.0135}_{-0.0099}$ \TTstrut\\
                 && Observed & $533$ & $10$ & $1$ & $0$  \TTstrut\\

\cline{2-7}

         & \multirow{4}{*}{5318  -- 5338 }
                 & Expected comb. bkg & $566.8^{+6.3}_{-5.8}$ & $6.90^{+0.61}_{-0.55}$ & $1.16^{+0.38}_{-0.34}$ & $0.109^{+0.079}_{-0.063}$ \TTstrut\\
                 && Expected peak. bkg & $0.052^{+0.023}_{-0.018}$ & $0.054^{+0.026}_{-0.019}$ & $0.052^{+0.024}_{-0.018}$ & $0.051^{+0.023}_{-0.018}$ \TTstrut\\
                 && Expected signal & $0.205^{+0.073}_{-0.074}$ & $0.114^{+0.040}_{-0.031}$ & $0.142^{+0.036}_{-0.025}$ & $0.142^{+0.042}_{-0.031}$ \TTstrut\\
                 && Observed & $525$ & $9$ & $0$ & $1$  \TTstrut\\

\cline{2-7}

         & \multirow{4}{*}{5338  -- 5358 }
                 & Expected comb. bkg & $558.2^{+6.1}_{-5.6}$ & $6.84^{+0.59}_{-0.54}$ & $1.14^{+0.37}_{-0.33}$ & $0.106^{+0.075}_{-0.060}$ \TTstrut\\
                 && Expected peak. bkg & $0.024^{+0.028}_{-0.012}$ & $0.025^{+0.026}_{-0.012}$ & $0.024^{+0.027}_{-0.012}$ & $0.025^{+0.025}_{-0.012}$ \TTstrut\\
                 && Expected signal & $0.38^{+0.14}_{-0.14}$ & $0.213^{+0.075}_{-0.058}$ & $0.267^{+0.065}_{-0.047}$ & $0.265^{+0.077}_{-0.058}$ \TTstrut\\
                 && Observed & $561$ & $6$ & $2$ & $1$  \TTstrut\\

\cline{2-7}

         & \multirow{4}{*}{5358  -- 5378 }
                 & Expected comb. bkg & $549.8^{+6.0}_{-5.4}$ & $6.77^{+0.57}_{-0.52}$ & $1.11^{+0.36}_{-0.32}$ & $0.103^{+0.073}_{-0.057}$ \TTstrut\\
                 && Expected peak. bkg & $0.0145^{+0.0220}_{-0.0091}$ & $0.0151^{+0.0230}_{-0.0091}$ & $0.0153^{+0.0232}_{-0.0098}$ & $0.015^{+0.023}_{-0.010}$ \TTstrut\\
                 && Expected signal & $0.38^{+0.14}_{-0.14}$ & $0.213^{+0.075}_{-0.057}$ & $0.267^{+0.065}_{-0.047}$ & $0.265^{+0.077}_{-0.057}$ \TTstrut\\
                 && Observed & $515$ & $7$ & $0$ & $0$  \TTstrut\\

\cline{2-7}

         & \multirow{4}{*}{5378  -- 5398 }
                 & Expected comb. bkg & $541.5^{+5.8}_{-5.3}$ & $6.71^{+0.55}_{-0.51}$ & $1.09^{+0.34}_{-0.31}$ & $0.101^{+0.070}_{-0.054}$ \TTstrut\\
                 && Expected peak. bkg & $0.0115^{+0.0175}_{-0.0086}$ & $0.0116^{+0.0177}_{-0.0090}$ & $0.0118^{+0.0179}_{-0.0090}$ & $0.0118^{+0.0179}_{-0.0088}$ \TTstrut\\
                 && Expected signal & $0.204^{+0.073}_{-0.074}$ & $0.114^{+0.040}_{-0.031}$ & $0.142^{+0.036}_{-0.026}$ & $0.141^{+0.042}_{-0.031}$ \TTstrut\\
                 && Observed & $547$ & $10$ & $1$ & $1$  \TTstrut\\

\cline{2-7}

         & \multirow{4}{*}{5398  -- 5418 }
                 & Expected comb. bkg & $533.4^{+5.7}_{-5.2}$ & $6.65^{+0.53}_{-0.49}$ & $1.07^{+0.34}_{-0.30}$ & $0.098^{+0.068}_{-0.051}$ \TTstrut\\
                 && Expected peak. bkg & $0.0089^{+0.0136}_{-0.0065}$ & $0.0088^{+0.0133}_{-0.0066}$ & $0.0091^{+0.0138}_{-0.0070}$ & $0.0090^{+0.0137}_{-0.0065}$ \TTstrut\\
                 && Expected signal & $0.058^{+0.024}_{-0.021}$ & $0.0323^{+0.0128}_{-0.0093}$ & $0.0407^{+0.0120}_{-0.0087}$ & $0.0402^{+0.0137}_{-0.0097}$ \TTstrut\\
                 && Observed & $501$ & $4$ & $1$ & $0$  \TTstrut\\
\hline
\end{tabular}

}
\end{center}
\label{tab:data_bsmm}
\end{table}

\begin{table}[t]
\tabcaption{Expected combinatorial background events, expected peaking (\Bhh) background events, 
expected \Bdmm signal events assuming the SM \BF, 
expected cross-feed events from \Bsmumu assuming the SM \BF
and observed events in the \Bdmumu search window.}
\begin{center}
\resizebox {15.9cm }{!}{%
\begin{tabular}{|l c c|c|c|c|c|}
\hline
& & & \multicolumn{4}{c|}{BDT\TTstrut\BBstrut}  \\
& & & 0. -- 0.25  & 0.25 -- 0.5 & 0.5 -- 0.75 & 0.75 -- 1. \\
\hline
\multirow{43}{*}{\rotatebox{90}{Invariant mass [\MeVcc]}}
         & \multirow{4}{*}{5212  -- 5232 }
                 & Expected comb. bkg & $614.2^{+7.5}_{-7.0}$ & $7.23^{+0.77}_{-0.68}$ & $1.31^{+0.46}_{-0.40}$ & $0.123^{+0.107}_{-0.072}$ \TTstrut\\
                 && Expected peak. bkg & $0.203^{+0.038}_{-0.034}$ & $0.206^{+0.038}_{-0.034}$ & $0.203^{+0.037}_{-0.034}$ & $0.205^{+0.038}_{-0.034}$ \TTstrut\\
                 && Cross-feed & $0.0056^{+0.0021}_{-0.0020}$ & $0.00312^{+0.00119}_{-0.00087}$ & $0.00391^{+0.00107}_{-0.00078}$ & $0.00387^{+0.00122}_{-0.00092}$ \TTstrut\\             
                 && Expected signal & $0.0070^{+0.0027}_{-0.0026}$ & $0.0039^{+0.0015}_{-0.0011}$ & $0.0049^{+0.0014}_{-0.0010}$ & $0.0048^{+0.0016}_{-0.0012}$ \TTstrut\\
                 && Observed & $554$ & $6$ & $0$ & $2$  \TTstrut\\

\cline{2-7}

         & \multirow{4}{*}{5232  -- 5252 }
                 & Expected comb. bkg & $605.0^{+7.2}_{-6.8}$ & $7.17^{+0.74}_{-0.65}$ & $1.29^{+0.44}_{-0.39}$ & $0.121^{+0.102}_{-0.072}$ \TTstrut\\
                 && Expected peak. bkg & $0.281^{+0.056}_{-0.049}$ & $0.279^{+0.056}_{-0.049}$ & $0.280^{+0.056}_{-0.049}$ & $0.280^{+0.058}_{-0.050}$ \TTstrut\\
                 && Cross-feed & $0.0071^{+0.0027}_{-0.0026}$ & $0.0039^{+0.0015}_{-0.0011}$ & $0.00496^{+0.00134}_{-0.00099}$ & $0.0049^{+0.0016}_{-0.0012}$ \TTstrut\\            
                 && Expected signal & $0.0241^{+0.0086}_{-0.0087}$ & $0.0135^{+0.0048}_{-0.0037}$ & $0.0169^{+0.0042}_{-0.0031}$ & $0.0167^{+0.0050}_{-0.0037}$ \TTstrut\\
&& Observed & $556$ & $4$ & $2$ & $1$  \TTstrut\\
\cline{2-7}

         & \multirow{4}{*}{5252  -- 5272 }
                 & Expected comb. bkg & $595.9^{+7.0}_{-6.5}$ & $7.10^{+0.71}_{-0.63}$ & $1.26^{+0.42}_{-0.37}$ & $0.119^{+0.097}_{-0.072}$ \TTstrut\\
                 && Expected peak. bkg & $0.323^{+0.075}_{-0.061}$ & $0.326^{+0.074}_{-0.061}$ & $0.324^{+0.072}_{-0.060}$ & $0.325^{+0.075}_{-0.062}$ \TTstrut\\
                 && Cross-feed & $0.0097^{+0.0036}_{-0.0035}$ & $0.0054^{+0.0021}_{-0.0015}$ & $0.0068^{+0.0018}_{-0.0013}$ & $0.0067^{+0.0021}_{-0.0016}$ \TTstrut\\
                 && Expected signal & $0.045^{+0.016}_{-0.016}$ & $0.0252^{+0.0088}_{-0.0067}$ & $0.0317^{+0.0077}_{-0.0057}$ & $0.0313^{+0.0093}_{-0.0068}$ \TTstrut\\
                 && Observed & $588$ & $11$ & $1$ & $0$  \TTstrut\\

\cline{2-7}

         & \multirow{4}{*}{5272  -- 5292 }
                 & Expected comb. bkg & $586.9^{+6.7}_{-6.3}$ & $7.04^{+0.68}_{-0.60}$ & $1.23^{+0.41}_{-0.36}$ & $0.117^{+0.092}_{-0.071}$ \TTstrut\\
                 && Expected peak. bkg & $0.252^{+0.058}_{-0.047}$ & $0.252^{+0.056}_{-0.046}$ & $0.253^{+0.059}_{-0.048}$ & $0.250^{+0.056}_{-0.046}$ \TTstrut\\
                 && Cross-feed & $0.0154^{+0.0058}_{-0.0055}$ & $0.0086^{+0.0033}_{-0.0024}$ & $0.0108^{+0.0029}_{-0.0021}$ & $0.0106^{+0.0033}_{-0.0025}$ \TTstrut\\
                 && Expected signal & $0.045^{+0.016}_{-0.016}$ & $0.0251^{+0.0089}_{-0.0067}$ & $0.0317^{+0.0077}_{-0.0057}$ & $0.0313^{+0.0092}_{-0.0069}$ \TTstrut\\
                 && Observed & $616$ & $5$ & $2$ & $1$  \TTstrut\\

\cline{2-7}

         & \multirow{4}{*}{5292  -- 5312 }
                 & Expected comb. bkg & $578.1^{+6.5}_{-6.1}$ & $6.98^{+0.66}_{-0.58}$ & $1.20^{+0.39}_{-0.35}$ & $0.114^{+0.087}_{-0.067}$ \TTstrut\\
                 && Expected peak. bkg & $0.124^{+0.023}_{-0.021}$ & $0.124^{+0.023}_{-0.021}$ & $0.123^{+0.023}_{-0.021}$ & $0.124^{+0.023}_{-0.021}$ \TTstrut\\
                 && Cross-feed & $0.038^{+0.015}_{-0.014}$ & $0.0214^{+0.0086}_{-0.0061}$ & $0.0270^{+0.0080}_{-0.0056}$ & $0.0266^{+0.0089}_{-0.0064}$ \TTstrut\\
                 && Expected signal & $0.0241^{+0.0086}_{-0.0087}$ & $0.0134^{+0.0048}_{-0.0036}$ & $0.0169^{+0.0042}_{-0.0030}$ & $0.0167^{+0.0050}_{-0.0037}$ \TTstrut\\
                 && Observed & $549$ & $7$ & $0$ & $0$  \TTstrut\\

\cline{2-7}

         & \multirow{4}{*}{5312  -- 5332 }
                 & Expected comb. bkg & $569.3^{+6.3}_{-5.9}$ & $6.92^{+0.63}_{-0.57}$ & $1.18^{+0.38}_{-0.34}$ & $0.111^{+0.083}_{-0.064}$ \TTstrut\\
                 && Expected peak. bkg & $0.047^{+0.023}_{-0.012}$ & $0.047^{+0.022}_{-0.012}$ & $0.047^{+0.021}_{-0.012}$ & $0.047^{+0.021}_{-0.012}$ \TTstrut\\
                 && Cross-feed & $0.149^{+0.055}_{-0.054}$ & $0.083^{+0.031}_{-0.022}$ & $0.104^{+0.027}_{-0.019}$ & $0.103^{+0.031}_{-0.023}$ \TTstrut\\        
                 && Expected signal & $0.0068^{+0.0028}_{-0.0026}$ & $0.0038^{+0.0015}_{-0.0011}$ & $0.0048^{+0.0014}_{-0.0010}$ & $0.0048^{+0.0016}_{-0.0012}$ \TTstrut\\
                 && Observed & $509$ & $10$ & $1$ & $1$  \TTstrut\\
\hline
\end{tabular}
}
\end{center}
\label{tab:data_bdmm}
\end{table}

\begin{table}[!htb]
\tabcaption{Expected and observed limits on the \Bsmumu branching fraction for 
the 2011 data and for the combination of 2010 and 2011 data. The expected limits are computed allowing 
the presence of \Bsmumu events according to the SM branching fraction.}
\label{tab:bs_results}
\begin{center}
\begin{tabular}{ccccc}
\toprule 
      &   & at 90\% CL & at 95\% CL & \CLb\\ 
\midrule
2011 & expected limit         & {$1.1 \times 10^{-8}$} & {$ 1.4 \times 10^{-8}$} &  \\ 

     & observed limit         &$1.3 \times 10^{-8}$ & $ 1.6 \times 10^{-8}$  &  0.95 \\ 
\midrule
2010+2011 & expected limit   & {$1.0 \times 10^{-8}$} & {$ 1.3 \times 10^{-8}$} &  \\ 

   & observed limit          &$1.2 \times 10^{-8}$ & $ 1.4 \times 10^{-8}$  &  0.93 \\ 

\bottomrule
\end{tabular}
\end{center}
\end{table}

\begin{table}[!htb]
\tabcaption{Expected and observed limits on the \Bdmumu branching fraction for 2011 data and for 
the combination of 2010 and 2011 data. The expected limits are computed in the background only hypothesis.}
\label{tab:bd_results}
\begin{center}
\begin{tabular}{ccccc}
\toprule 
 &    & at 90\% CL & at 95\% CL & \CLb \\ 
\midrule 
2011 & expected limit          &  {$2.5 \times 10^{-9}$} & {$ 3.2 \times 10^{-9}$} & \\ 

 & observed limit         & $3.0\times 10^{-9}$ & $ 3.6 \times 10^{-9}$  & 0.68\\ 
\midrule 

2010+2011 & expected limit  &  {$2.4 \times 10^{-9}$} & {$ 3.0 \times 10^{-9}$} & \\ 

 & observed limit        & $2.6\times 10^{-9}$ & $ 3.2 \times 10^{-9}$  & 0.61\\ 
\bottomrule
\end{tabular}
\end{center}
\end{table}

\section{Conclusions}
\label{sec:conclusions}

With 0.37\invfb of integrated luminosity, a search for the 
rare decays \Bsmumu and \Bdmumu has been performed and sensitivities 
better than the existing limits have been obtained.
The observed events in the \Bs and in the \Bd mass windows are compatible with the background expectations 
at 5\% and 32\% confidence level, respectively.
For the \Bsmumu decay, the probability that the observed events are compatible with the 
sum of expected background events 
and signal events according to the SM rate is
33\%.
The upper limits for the branching fractions are evaluated to be
\begin{eqnarray}
\BRof{\Bsmm} &<& 1.3 \,(1.6) \times 10^{-8}~{\rm at}~90\,\% \,(95\,\%)~{\rm CL},  \nonumber \\
\BRof{\Bdmm} &<& 3.0 \,(3.6) \times 10^{-9}~{\rm at}~90\,\% \,(95\,\%)~{\rm CL}.  \nonumber
\end{eqnarray} 

\noindent
The \BRof \Bsmumu and \BRof \Bdmumu upper limits have 
been combined with those  published previously by LHCb  \cite{LHCb_paper} and the results are
\begin{eqnarray}
\BRof{\Bsmm} (2010+2011) &<& 1.2 \,(1.4) \times 10^{-8}~{\rm at}~90\,\% \,(95\,\%)~{\rm CL},  \nonumber \\
\BRof{\Bdmm} (2010+2011) &<& 2.6 \,(3.2) \times 10^{-9}~{\rm at}~90\,\% \,(95\,\%)~{\rm CL}.  \nonumber
\end{eqnarray} 
The above 90\% (95\%) \CL upper limits are still about 3.8 (4.4) times the SM branching fractions 
for the \Bs and 26 (32) times for the \Bd.
These results represent the best upper limits to date.

\clearpage
\section*{Acknowledgements}

\noindent We express our gratitude to our colleagues in the CERN accelerator
departments for the excellent performance of the LHC. We thank the
technical and administrative staff at CERN and at the LHCb institutes,
and acknowledge support from the National Agencies: CAPES, CNPq,
FAPERJ and FINEP (Brazil); CERN; NSFC (China); CNRS/IN2P3 (France);
BMBF, DFG, HGF and MPG (Germany); SFI (Ireland); INFN (Italy); FOM and
NWO (The Netherlands); SCSR (Poland); ANCS (Romania); MinES of Russia and
Rosatom (Russia); MICINN, XuntaGal and GENCAT (Spain); SNSF and SER
(Switzerland); NAS Ukraine (Ukraine); STFC (United Kingdom); NSF
(USA). We also acknowledge the support received from the ERC under FP7
and the Region Auvergne.

\bibliographystyle{LHCb}
\bibliography{paper025}

\end{document}